\documentclass[times,twocolumn,final]{elsarticle}

\usepackage{medima}
\usepackage{framed,multirow}

\usepackage{amsmath,amssymb,amsfonts}
\usepackage{latexsym}

\usepackage{url}
\usepackage{xcolor}
\usepackage{hyperref}

\usepackage{lineno}
\usepackage{soul}
\usepackage{algorithmic}
\usepackage{graphicx}
\usepackage{textcomp}
\usepackage{textgreek}
\usepackage{hyperref}
\usepackage{makecell}
\usepackage{threeparttable}
\usepackage[utf8]{inputenc}
\usepackage[normalem]{ulem}
\useunder{\uline}{\ul}{}

\definecolor{newcolor}{rgb}{.8,.349,.1}

\graphicspath{ {./images/} }
\newcommand{\gh}[1]{\textcolor{black}{{#1}}}
\newcommand{\ghr}[1]{\textcolor{black}{{#1}}}

\newcommand{
\beginappendix}{%
\setcounter{table}{0}
\renewcommand{\thetable}{S\arabic{table}}%
\setcounter{figure}{0}
\renewcommand{\thefigure}{S\arabic{figure}}%
}

\def\endabstract{\egroup}

\journal{NeuroImage}

\begin{document}

\verso{Hao Guan \textit{et~al.}}

\begin{frontmatter}

\title{MRI-based Alzheimer's disease prediction via distilling the knowledge in multi-modal data}

\author[addr1]{Hao Guan}
\author[addr1]{Chaoyue Wang\corref{cor1}}
\author[addr1,addr2]{Dacheng Tao\corref{cor1}}

\cortext[cor1]{Corresponding author.\\
\indent\;\; \textit{E-mail address}: \rm\{chaoyue.wang, dacheng.tao\}@sydney.edu.au.}

\address[addr1]{ School of Computer Science, The University of Sydney, Australia}
\address[addr2]{ JD Explore Academy, China}



\begin{abstract}
Mild cognitive impairment (MCI) conversion prediction, \textit{i.e.,} identifying MCI patients of high risks converting to Alzheimer's disease (AD), is essential for preventing or slowing the progression of AD. Although previous studies have shown that the fusion of multi-modal data can effectively improve the prediction accuracy, their applications are largely restricted by the limited availability or high cost of multi-modal data. Building an effective prediction model using only magnetic resonance imaging (MRI) remains a challenging research topic. 
In this work, we propose a multi-modal multi-instance distillation scheme, which aims to distill the knowledge learned from multi-modal data to an MRI-based network for MCI conversion prediction. In contrast to existing distillation algorithms, the proposed multi-instance probabilities demonstrate a superior capability of representing the complicated atrophy distributions, and can guide the MRI-based network to better explore the input MRI. To our best knowledge, this is the first study that attempts to improve an MRI-based prediction model by leveraging extra supervision distilled from multi-modal information. Experiments demonstrate the advantage of our framework, suggesting its potentials in the data-limited clinical settings.
\end{abstract}


\end{frontmatter}


\section{Introduction}\label{sec1}



Mild cognitive impairment (MCI) is a transitional stage between cognitively normal and dementia \citep{Petersen2004}. In each year, around 10\%–15\% of MCI patients convert to AD \citep{Mitchell2008}, while the other MCI patients tend to remain stable, develop other forms of dementia, or even revert to normal cognition \citep{Pandya2016}. The therapeutic interventions are thought to be feasible for MCI before it progressing to the irreversible AD \citep{Carretti2013,Sherman2017,Yao2020}. Therefore, it is of great importance to identify the MCI patients who have high risk of converting to AD, allowing early treatments to prevent or delay the progression of the disease. 

The AD process is associated with plaques and neurofibrillary tangles in the brain \citep{Ballard2011}. AD-related neuropathology can be identified via a series of factors several years before AD clinical manifestation \citep{Reiman2010,Ewers2011,Hampel2018,Teipel2018}. For example, MRI detects grey matter volume changes caused by gross neuronal loss and atrophy \citep{Ewers2011}. And the E4 variant of apolipoprotein E (APOe4) is the main susceptibility gene for Alzheimer's disease \citep{Corder1993,Roses1996,Genin2011}. Multi-modal data, including imaging and non-imaging clinical data, provide rich and complementary information about the disease. Previous studies have identified some genetic basis of phenotypic neuroimaging markers \citep{Shen2010}, and brain imaging has revealed correlations between brain structure and cognition in AD and MCI \citep{Braskie2013}. Based on these findings, extensive computer-aided diagnosis (CAD) studies have demonstrated that employing multi-modal data can largely improve the performance of AD diagnosis and MCI conversion prediction \citep{Liu2017,Zhou2019,Zhu2019,Moradi2015,Spasov2019,Qiu2020}. However, it is difficult to meet the setting of multi-modal data in practice, because the medical data of some certain modalities are of high cost or limited availability. 
The requirement of multi-modal data restricts the applicability of the above multi-modal based CAD methods \citep{Liu2017,Zhou2019,Zhu2019,Moradi2015,Spasov2019,Qiu2020}.

MRI is non-invasive without using ionizing radiation. The comparatively low cost makes MRI be widely applied in the clinical practice. In recent years, MRI-based CAD methods have achieved promising performance in AD diagnosis \citep{Bron2015,Liu2016,Liu2018,Lian2018}. However, since the disease-related patterns in MCI brains are more subtle than those in AD brains \citep{Driscoll2009}, MCI conversion prediction appears to be a more challenging task \citep{Liu2016,Liu2018,Lian2018,Pan2020}. Compared with multi-modal based CAD methods \citep{Moradi2015,Spasov2019,Pan2020,Lin2018}, existing MRI-based methods usually perform worse in MCI conversion prediction \citep{Liu2016,Liu2018,Lian2018}. This suggests that when merely using MRI data, extracting disease-related features from MCI brain images is difficult. To achieve better prediction performance, a CAD algorithm has to extract fine-grained features that contain more information about the disease patterns and less disease-irrelevant information.

Knowledge distillation is typically used to transfer knowledge from a larger teacher network to a small student network \citep{Hinton2015,Ruffy2019,Gou2020}. Similarly, the knowledge learned by a teacher from privileged data can also be transferred to a student without accessing that data \citep{Vapnik2015,LopezPaz2016}. For a classification task, the relative probabilities of incorrect classes provide the knowledge that guides the student network to generalize \citep{Hinton2015}. By minimizing the Kullback Leibler (KL) divergence between the outputs of a teacher and a student, the student receives extra supervision beyond the conventional training loss. However, for a binary classification problem like MCI conversion prediction, the output probabilities of a teacher network are not sufficiently informative for a student to mimic \citep{Shen2016}, since the relationship between classes is not leveraged \citep{Tang2020}. Meanwhile, the gap between data of different modalities raises the challenge of how to transfer knowledge from a multi-modal teacher network to an MRI-based student network.

\begin{figure}[t!]
\centerline{\includegraphics[width=\columnwidth]{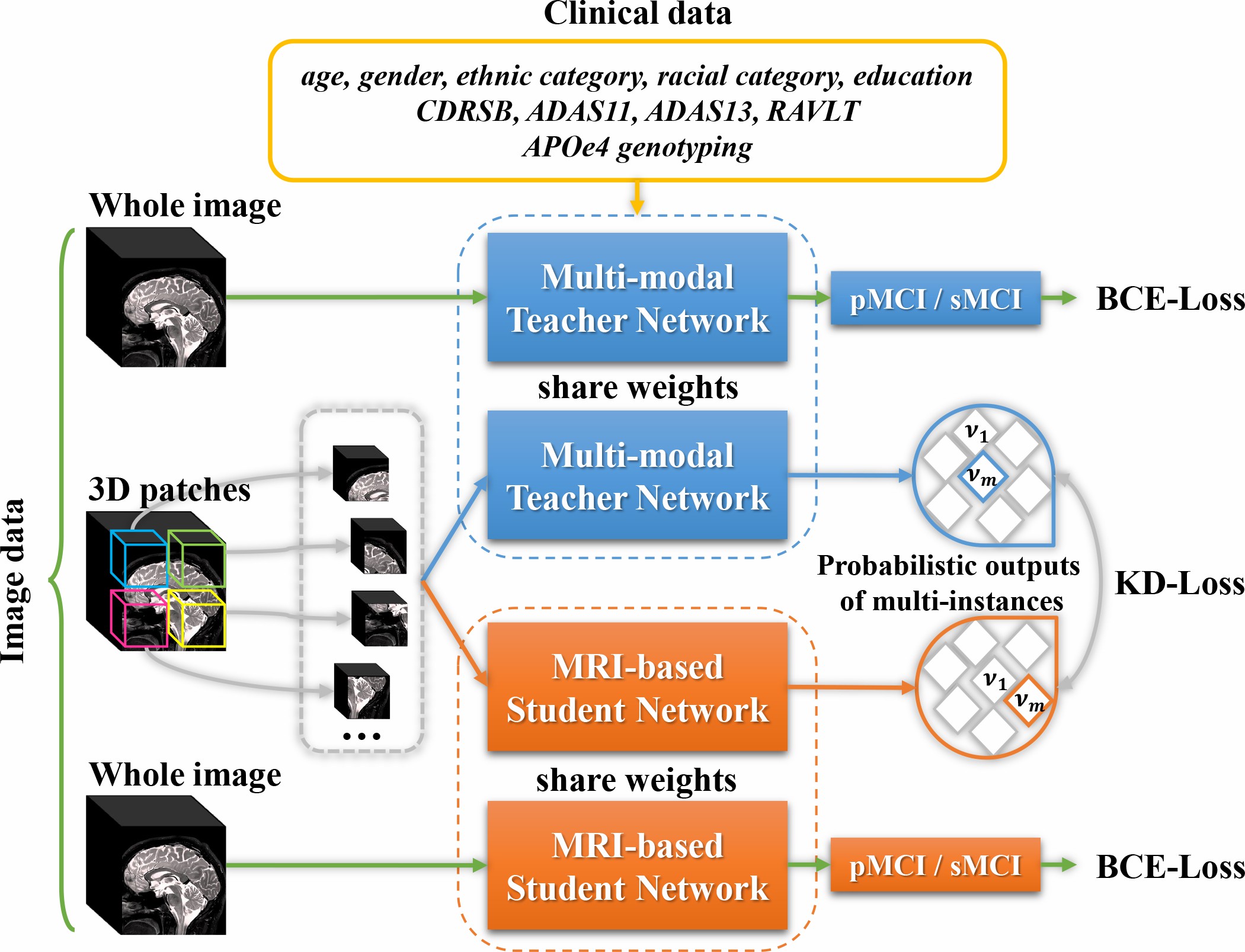}}\vspace{-0.1 cm}
\caption{The schematic of the proposed framework. The teacher network takes MRI and clinical data as input. The student network is only fed with MRI images. Knowledge is encoded and transferred in the form of probabilistic outputs of multi-instances. Abbreviations: APOe4, Apolipoprotein E; CDRSB, Clinical Dementia Rating Sum of Boxes; ADAS, Alzheimer's disease Assessment Scale; RAVLT, Ray Auditory Verbal Learning Test; BCE-Loss: binary cross-entropy loss; KD-Loss: knowledge distillation loss.} 
\label{fig1}
\end{figure}

In this paper, instead of directly using the multi-modal data for MCI conversion prediction, we propose a novel multi-modal multi-instance distillation (M3ID) method for achieving better MRI-based prediction. The schematic diagram of our framework is shown in Fig. \ref{fig1}. We perform knowledge transfer between a multi-modal teacher network (using multi-modal data as input) and a single-modal student network (using only MRI as input). Specifically, multi-modal data include MRI and non-imaging clinical data.
Different from existing distillation methods, we devise a probabilities matching scheme regarding multi-instances: we use the KL-Divergence to reduce the distance between the teacher's and the student's probabilistic estimates of multi-instances. Specifically, we partition each MRI scan into multiple 3D image patches, which make individual instances. The probability estimates of multi-instances are obtained by feeding the networks with image patches, and used as the extra supervision for the student network. Thus, the student is guided to mimic the teacher's modeling of subtle disease patterns and achieve better prediction performance. Overall, we forge a connection between \emph{knowledge distillation} and \emph{multi-instance learning}. Based on this connection, we are able to leverage knowledge learned from multi-modal data to improve the MRI-based model's performance. We conducted experiments on three datasets for MCI conversion prediction. Experimental results demonstrate our framework's superior prediction performance. 

Our contributions are summarized as follows:
\begin{enumerate}
    \item This is the first study that aims to improve the MRI-based prediction of MCI conversion by leveraging extra supervision distilled from multi-modal information;
    \item Our work facilitates the clinical practices where multi-modal data are available for model training and only MRI data are used for testing;
    \item We present a multi-instance-based knowledge distillation scheme, and the experiments on three independent datasets demonstrate the advantage our proposed scheme;
    \item Our framework can generate heat-maps on the basis of image patches for visual interpretation.
\end{enumerate}

The rest of the paper is organized as follows. In Section \ref{rw}, we review related works. In Section \ref{ms}, we introduce the studied datasets, and our new method. In Section \ref{ea}, we evaluate the proposed method. In Section \ref{ds}, we compare our method with previous methods and discuss the limitations of the current study. In Section \ref{co}, we conclude this work.

\section{Related Work}\label{rw}

\subsection{Knowledge Distillation}
Knowledge distillation aims to train a student network under the guidance of a trained teacher network. With additional knowledge beyond the usual data supervision (\textit{e.g.,} annotations), the student can be better optimized. The pioneering knowledge distillation work proposed by \citet{Hinton2015} minimizes the Kullback-Leibler (KL) divergence between output probabilities of a teacher and a student network. The relative probabilities of incorrect classes convey useful information about how a teacher tends to generalize. This distillation strategy is simple yet effective, and has been used in various applications \citep{Gou2020}. Some other methods transfer knowledge by representation mimicking. FitNets \citep{Romero2014} matches the full feature maps between a teacher and a student. Attention transfer \citep{Zagoruyko2017} attempts to match the spatial attention maps. Apart from directly matching a teacher's representation, similarity preserving \citep{Tung2019} encourages a student to preserve the pairwise activation similarities derived by a teacher network. Representation-based distillation methods bypass the problem of few output classes conveying limited knowledge. Although these representation-based methods have achieved reasonable performance, their effectiveness has not been validated when representations are generated from data of very different modalities.

Cross-modal knowledge transfer is an arising research topic of knowledge distillation. Given a teacher network trained on the source modality, the teacher's knowledge is transferred to a student network applied on a different modality \citep{Gupta2016,Passalis2018}. Knowledge transfer among different modalities enables effective utilizing of complementary information provided by multi-modal data. \citet{Do2019} propose a visual question answering method, in which a teacher model learns trilinear interactions from image-question-answer inputs, then transfers knowledge to a student model that learns bilinear interactions from image-question inputs. Cross-modal distillation fits the scenarios where certain modalities of data or labels are not available during the training or test stage. \citet{Dou2020} leverage cross-modal distillation to address the unpaired multi-modal segmentation task. \citet{Li2020} exploit the modality-shared knowledge to facilitate image segmentation of the target-modality. Based on the above studies, we investigate the knowledge transfer from multi-modal data to a single-modal disease prediction model.

\subsection{Deep Learning in AD Diagnosis}
Recently, many deep learning-based methods are proposed for AD diagnosis. MRI has the advantages of non-invasively capturing disease-related pathological patterns and common availability. Thus, a relatively large amount of MRI data can be used for training deep learning-based AD diagnosis models. \citet{Liu2018} exploit anatomical landmarks to develop an MRI-based deep learning model for AD classification and MCI conversion prediction. \citet{Lian2018} propose an AD diagnosis network with integration of discriminative atrophy localization, feature extraction and classifier construction. \citet{Lee2019} aim to build an interpretable diagnosis model by extracting regional abnormality representation from MRI. Despite all these efforts, multi-modal methods have always demonstrated a better diagnostic performance than the single-modal MRI-based methods. \citet{Spasov2019} devise a light-weight network that takes a combination of MRI, cognitive measurements, genetic, and demographic data as input, and achieved impressive results on MCI conversion prediction. \citet{Qiu2020} utilize multi-modal inputs of MRI, age, gender, and Mini-Mental State Examination score for training a deep learning model, and demonstrate the model's effectiveness of AD diagnosis on several datasets. \citet{Pan2020} propose a diagnosis framework utilizing MRI and PET data, and achieve promising results. However, these multi-modal methods seem to underutilize the MRI data, since the performance with the data other than MRI is already high. In practice, multi-modal medical data are hard to get, leaving room for further improvement using only MRI data.

\subsection{Multi-instance learning}
Multi-instance learning (MIL) tries to solve a weakly supervised learning problem that each labelled bag in a training set contains multiple unlabelled instances \citep{Amores2013,Carbonneau2018}. The goal of MIL is to predict the class labels of unseen bags. Under the MIL setting, a positive bag contains at least one positive instance, while all instances in a negative bag are negative. Compared with fully supervised learning methods, MIL reduces the annotation cost and fits many real-life applications. For example, by leveraging weakly labelled training videos, a deep multiple instance ranking framework is proposed for anomaly detection in surveillance videos \citep{Sultani2018}. A multi-instance deep learning method is developed to discover discriminative local anatomies for bodypart recognition \citep{Yan2016}. In the domain of medical image analysis, obtaining pixel-level annotation costs very much. With MIL and the weak supervision of per image diagnosis, a large quantity of medical images can be used for model training \citep{Kandemir2015}. An attention-based MIL method presented in \citep{Ilse2018} achieves promising results on cancer diagnosis via histology image. A landmark-based deep MIL framework use MRI data \citep{Liu2018} for brain disease diagnosis. These studies suggests that multi-instance representations leverage rich information derived from data. In the present study, instead of using MIL for producing the final diagnosis, we leverage multi-instance probabilities as the medium for knowledge distillation.

\section{Material and methods}\label{ms}
In this section, we first introduce the datasets and MRI pre-processing pipeline used in this study (Section \ref{ms1}). Then we present the proposed multi-modal multi-instance distillation scheme including the mathematical formulation of model (Section \ref{ms2}), the network architectures of the proposed multi-modal teacher and the MRI-based student (Section \ref{ms3}), the novel multi-instance distillation method (Section \ref{ms4}), and the distillation loss and implementation details (Section \ref{ms5}-\ref{ms6}).

\subsection{Datasets and Image Pre-processing}\label{ms1}
We collected data from three public datasets, ADNI-1, ADNI-2 \citep{Jack2008}, and the Australian Imaging, Biomarkers and Lifestyle (AIBL) database \citep{Ellis2009} \footnote[1]{
Data used in this paper were obtained from the Alzheimer's Disease Neuroimaging Initiative (ADNI) dataset (\url{adni.loni.usc.edu}) and the Australian Imaging Biomarkers and Lifestyle Study of Aging (AIBL) database (\url{www.aibl.csiro.au}). As such, the investigators within the ADNI and AIBL contributed to the design and implementation of ADNI and AIBL and/or provided data but did not participate in analysis or writing of this report. A complete listing of ADNI investigators can be found at: \href{http://adni.loni.usc.edu/wp-content/uploads/how_to_apply/ADNI_Acknowledgement_List.pdf}{online}}. \ghr{The ADNI-3 dataset was not used because it hadn't release long enough follow-ups.} It is worth noting that all subjects in ADNI-2 were newly enrolled, so there are no subjects appeared in both ADNI datasets. Only baseline data were used in this study. 

All T1-weighted images were pre-processed using FSL's anatomical processing pipeline \footnote[2]{\url{http://fsl.fmrib.ox.ac.uk/fsl/fslwiki/fsl_anat}} \citep{Jenkinson2012}. First, all images were reoriented to the standard space (Montreal Neurological Institute, MNI) and automatically cropped before bias-field correction. Next, the corrected images were registered non-linearly using the MNI T1 Template (MNI152\_T1\_1mm) \citep{Grabner2006}. Then, the skull was stripped using the FNIRT-based approach \citep{Jenkinson2012}, and the brain-extracted images were regenerated to a standard space of size $182\times 218\times 182$. Processed images had an identical spatial resolution ($1mm\times 1mm \times 1mm$). To further reduce the redundant computational cost, each image was cropped to $144\times 176\times 144$ (see Fig. \ref{figs7}). The 3D images were standardized to zero mean and unit variance before feeding into the network. Pre-processed images were manually checked, and the images with insufficient stereotaxic registration and insufficient skull stripping were excluded.

Clinical data employed in this study consisted of 13 scalar variables: age, gender, ethnic category, racial category, education years, Clinical Dementia Rating Sum of Boxes (1 variable), Alzheimer's disease assessment scale (2 variables: ADAS11, ADAS13), episodic memory evaluations in the Rey Auditory Verbal Learning Test (4 variables: immediate, learning, forgetting, percent forgetting), as well as APOe4 genotyping (1 variable). All clinical data used in this study were from baseline assessments. The clinical data were standardized to zero mean and unit variance before feeding into the network. The mean and standard deviation on the training set were used to apply the standardization on the validation and test set. The detailed clinical information of ADNI subjects was reported in Table \ref{tabs1} of the \emph{Appendix}.

\begin{table}[t!]
\centering
\caption{Demographic characteristics of the studied subjects.}\label{tab1}
\setlength{\tabcolsep}{3pt}
\begin{threeparttable}
\begin{tabular}{
p{30pt}<{\centering} p{35pt}<{\centering} p{35pt}<{\centering} p{50pt}<{\centering} p{60pt}<{\centering}
}
\hline
Dataset & Category & No. of subjects & Age Range & Females/Males \\
\hline
\multirow{4}{*}{ADNI-1} 
& sMCI  & 100 & 57.8-87.9 & 34/66    \\
& pMCI  & 164 & 55.2-88.3 & 66/98    \\
& NC    & 226 & 59.9-89.6 & 109/117  \\
& AD    & 179 & 55.1-90.9 & 87/92    \\
\hline
\multirow{4}{*}{ADNI-2} 
& sMCI  & 115 & 55.0-88.6 & 48/67    \\
& pMCI  & 76  & 56.5-85.9 & 36/40    \\
& NC    & 186 & 56.2-89.0 & 96/90    \\
& AD    & 140 & 55.6-90.3 & 56/84    \\
\hline
\end{tabular}
\end{threeparttable}
\end{table}

ADNI-1: The baseline ADNI-1 dataset used in this study consisted of 1.5T T1-weighted MRI images scanned from 669 subjects. According to standard clinical criteria, subjects were divided into three groups: normal control (NC), MCI, and AD. According to whether MCI subjects converted to AD within 36 months, MCI subjects were further categorized into: 1) sMCI (stable MCI), if diagnosis was MCI at all available time points (0-96 months), and had diagnosis records for at least 36 months; 2) pMCI (progressive MCI), if diagnosis was MCI at baseline but the conversion to AD was reported after baseline within 3 years, and without a reversion to MCI or NC at any available follow-up (0-96 months). The other MCI subjects who had the last diagnosis reported after baseline within 2 years, or had unstable diagnoses (reversion, or conversion after 3 years) were not included in this study. The dataset contained 100 sMCI, 164 pMCI, 226 NC, and 179 AD subjects The demographic information of the ADNI-1 subjects is presented in Table \ref{tab1}. 

ADNI-2: The baseline ADNI-2 dataset used in this study contained 3T T1-weighted MRI images scanned from 517 subjects. The same clinical criteria as in ADNI-1 were used to separate the subjects into 115 sMCI, 76 pMCI, 186 NC, and 140 AD subjects (see Table \ref{tab1}). 

AIBL: The AIBL dataset used in this study contained 3T T1-weighted MRI images scanned from 37 subjects, which consists of 25 sMCI and 12 pMCI. The demographic characters of the AIBL subjects were presented in Table \ref{tabs1} of the \emph{Appendix}.

\subsection{Mathematical formulation of model}\label{ms2}
Let $\{(\boldsymbol{X}_n, \boldsymbol{y}_n)\}^N_{n=1}$ be a training set containing N samples, where $\boldsymbol{X}_n$ denotes input data from the $n^{th}$ subject, and $\boldsymbol{y}_n$ denotes the corresponding class label (0 for sMCI, 1 for pMCI). The MCI conversion prediction task is formed as a classification problem, which can be formulated by estimating conditional probabilities of the target variable $\boldsymbol{y}$, given the input data $\boldsymbol{X}$:

\begin{equation}\label{eq1}
p(\boldsymbol{y}|\boldsymbol{X}; \theta),
\end{equation}
where $\theta$ represents the parameters of the neural network. The goal is to learn the network parameter so that the true conditional probabilities can be approximated. 

\subsection{Multi-modal Teacher \& Single-Modal Student}\label{ms3}
Given the scarcity and high cost of multi-modal data, we propose to improve an MRI-based prediction model by leveraging knowledge distillation. The complementary knowledge provided by multi-modal data is extracted by a teacher network. Then knowledge distillation is performed between the teacher network and a student network. After training, the student network is expected to effectively make disease prediction with only MRI data. Our prediction framework consists of a multi-modal attention network as the teacher, and an MRI-based network as the student. The details of the network architectures are given as follows.

\subsubsection{Student Network}
\begin{figure}[t!]
\centerline{\includegraphics[width=\columnwidth]{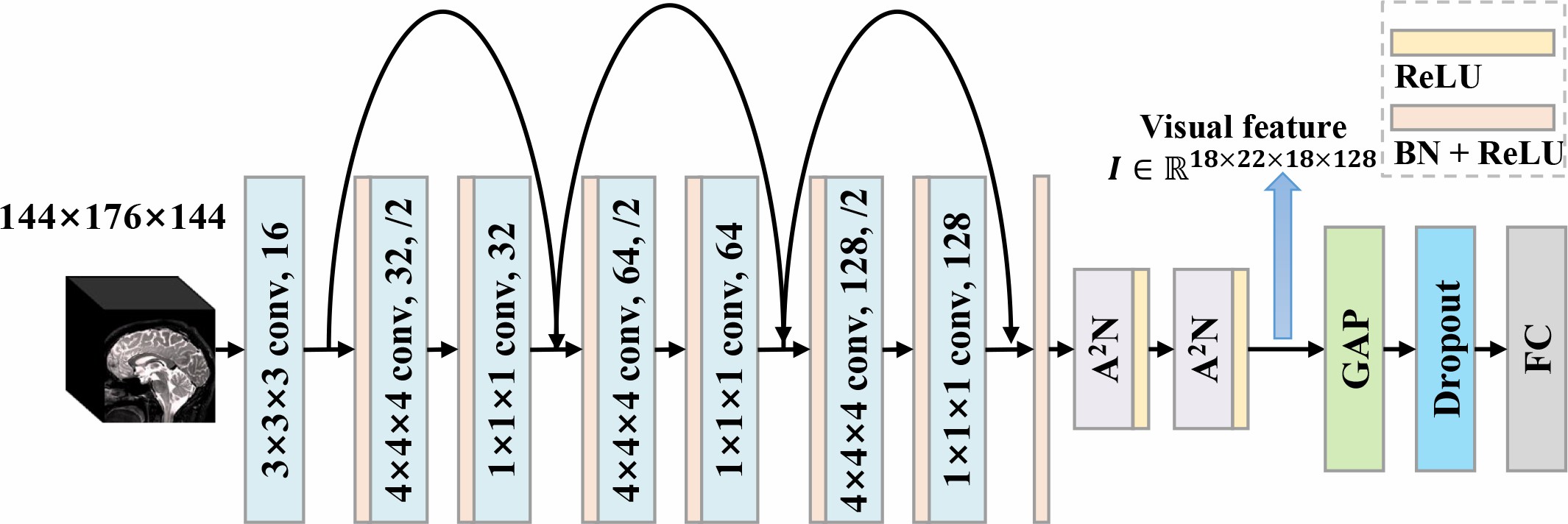}}
\caption{Architecture of the MRI-based student network.}\label{fig2}
\end{figure}

The student network is based on ResNet \citep{He2016} using residual block proposed in \citep{He2016a}. The network consists of one root convolutional layer and three residual blocks (Fig. \ref{fig2}). Feature down-sampling is achieved via strided convolutions. After three residual blocks, the feature map is down-sampled for 8 times. Then two double attention block ($A^{2}N$) \citep{Chen2018} are cascaded on top of the network to efficiently model the long-range inter-dependencies \citep{Burkov2018}. In the task of MRI-based MCI conversion prediction, the long-range inter-dependencies could be the spatial relationships between different areas of the brain. The following layers consist of a global average pooling layer, a dropout layer, and two fully connected layers. As shown in Fig. \ref{fig2}, the MRI-based student network takes a whole MRI scan as input, and aims to predict whether an MCI patient would convert to AD. Although existing MRI-based methods have achieved reasonable prediction results, there is still a big performance gap between them and multi-modal methods. Next we introduce a teacher network that uses multi-modal data.

\subsubsection{Teacher Network}
\begin{figure}[t!]
\centerline{\includegraphics[width=\columnwidth]{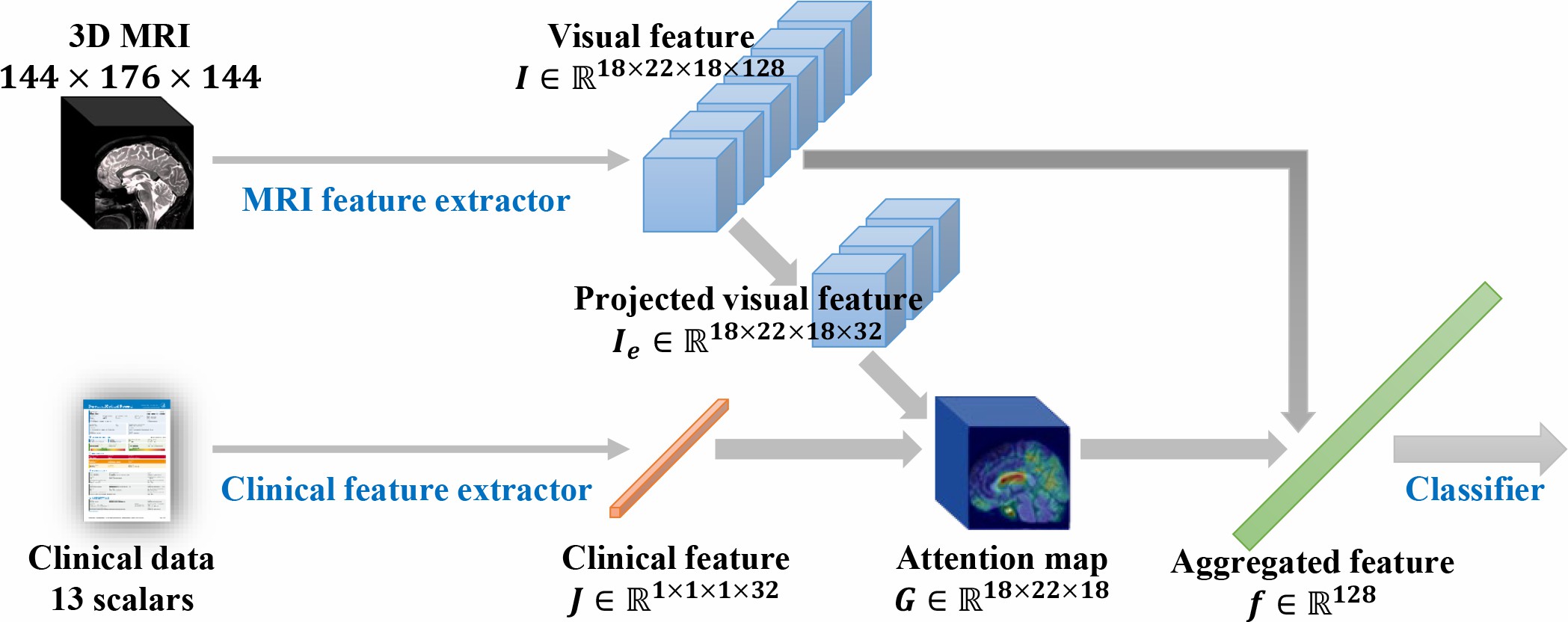}}
\caption{Diagram of the multi-modal attention network, \textit{i.e.,} the teacher network. The visual features are extracted and projected to the clinical space. The learned clinical features and the projected visual features are used to generate the attention maps. Then the visual features are weightedly averaged using the attention maps to generate the aggregated features.}\label{fig3}
\end{figure}

The teacher network utilizes an MRI feature extractor with the same architecture as the student network. In addition, a clinical feature extractor and a multi-modal attention module are introduced to build the whole teacher network, \textit{i.e.,} a multi-modal attention network. The learned clinical features are utilized to guide the generation of attention maps. Next, the learned attention maps are used to weightedly average the visual features and generate the aggregated features. The network architecture is inspired by \citep{You2019}. The diagram of the teacher network is shown in Fig. \ref{fig3}. The detailed definitions are given as follows.

The extracted visual features \gh{$\boldsymbol{I}$} are with spatial resolution of $H \times W \times D$ and \textit{C} channels. Firstly, we use two 1$\times$1$\times$1 convolutions to project the visual features to the clinical space. The projected features are denoted as \gh{$\boldsymbol{I}_{e} \in \mathbb{R}^{H\times W\times D \times E}$}, where \textit{E} represents the dimension of the clinical features. Following \citep{Spasov2019,Qiu2020}, the clinical data are passed through a sequence of two fully-connected layers to produce clinical features \gh{$\boldsymbol{J} \in \mathbb{R}^{E}$}. The multi-modal attention map is generated by computing the cosine similarity between the projected visual feature and \gh{$\boldsymbol{I}_{e}^{l}$} and clinical feature \gh{$\boldsymbol{J}$} at location \gh{$l$}.
\gh{
\begin{equation}\label{eq2} 
g^{l} = ReLU \frac{\boldsymbol{{I}_{e}^{l\boldsymbol{T}}} \boldsymbol{J}}
{\left\| \boldsymbol{{I}_{e}^{l}} \right\| \left\| \boldsymbol{J} \right\|}.
\end{equation}
}

The attention map $\boldsymbol{G}=[g^{1}, ... , g^{HWD}]$ is normalized by dividing the sum of attentive values across all space locations. The normalized attention map $\boldsymbol{U}=[u^{1}, ... , u^{HWD}]$ is used to combine visual features, \textit{i.e.,}
\gh{
\begin{equation}\label{eq3} 
\boldsymbol{f} = \sum_{l=1}^{HWD} u^{l} \textbf{I}^{l}.
\end{equation}
}

The attention-weighted features \gh{$\boldsymbol{f}$} are passed to two fully-connected layers for generating predictions. The teacher network is able to extract the complementary information and cross-modal dependencies from multi-modal data for better prediction. We choose to use the multi-modal attention network as the teacher network, because it is light-weight and effective, and has an architecture similar to the student network. The teacher network is a combination of the student network and the multi-modal attention module, which fuses multi-modal features and helps the teacher network achieve promising performance. The multi-modal attention network is designed to be similar to the student network, since the architecture difference between student and teacher network might degrade the performance of knowledge transfer \citep{Gou2020}.

\subsection{Multi-instance Knowledge Distillation}\label{ms4}
As aforementioned, the basic approach of knowledge distillation is to enforce a student to match the teacher's probabilistic outputs \citep{Hinton2015}. However, a binary classification model's outputs are not sufficiently informative to carry useful knowledge. A previous study has demonstrated that directly matching output probabilities does not offer enough guidance to a student with two output classes \citep{Shen2016}. To overcome this problem, we leverage multi-instance probabilities as a medium for knowledge distillation. A prediction network trained in a supervised manner is expected to learn a mapping from an input MRI to an output diagnostic label. When the network takes MRI patches as input, the outputs can be considered as the disease scores, which reflect the regional severities of the disease. The disease scores contain the information of how the network encodes the local regions.

Here, we employ patch-level disease scores to perform knowledge distillation. Specifically, we first partition the MRI into $M=N\times N\times N$ patches of size $x\times y\times z$, where \textit{x}, \textit{y} and \textit{z} are the coronal, sagittal, and axial sizes of the patches, respectively. The patches partitioned from the image $X$ make the individual instances in a bag. A network generates a disease score $\nu$ ranging from 0 to 1 for each MRI patch (\textit{i.e.,} instance). Then, softmax function is applied over a bag of disease scores $(\nu^1, \nu^2, ... , \nu^M)$ to produce the probabilistic estimates of multi-instances. The probability of instance \textit{m} of image $X$ given by the student and teacher are computed as:
\begin{equation}\label{eq4}
p^{m}_{S} = \frac{exp(\nu^{m}_{S})}{\sum^{M}_{m=1}exp(\nu^{m}_{S})} 
,\quad
p^{m}_{T} = \frac{exp(\nu^{m}_{T})}{\sum^{M}_{m=1}exp(\nu^{m}_{T})},
\end{equation}
where $p^{m}_{S}$ denotes the student's prediction, while $p^{m}_{T}$ denotes the teacher's prediction. A bag of disease scores represent the disease-related status distributed across the brain. With softmax activation, the multi-instance probabilities can convey useful information about the relationship between patches. When enforcing the student to mimic the teacher's probabilistic estimates of multi-instances, the teacher guides the student to better localize and grasp the disease-related information within MRI data. Intuitively, the proposed distillation method works like data augmentation, while the targets of patches are provided by a teacher rather than the human-annotated ground truth.

\subsection{Knowledge Transfer and Overall Loss Function}\label{ms5}
In our framework, the teacher's knowledge is encoded in the form of probabilities of multi-instances. We transfer the knowledge by minimizing the KL divergence between the two networks' probabilistic estimates of multi-instances, $\boldsymbol{p}_{T}$ and $\boldsymbol{p}_{S}$. The knowledge distillation (KD) loss is defined as follows:
\begin{equation}\label{eq5} 
\mathcal{L}_{KD} = D_{KL}(\boldsymbol{p}_{T} \lVert \boldsymbol{p}_{S})
=\sum^{N}_{i=1}\sum^{M}_{m=1}p^{m}_{T}\log{\frac{p^{m}_{T}}{p^{m}_{S}}}.
\end{equation}

In addition, a binary cross-entropy (BCE) loss is computed using the whole MRI as input and the corresponding conversion label as the target:
\begin{equation}\label{eq6}
\begin{aligned}
\mathcal{L}_{BCE}=-\frac{1}{N}\sum^{N}_{n=1}\boldsymbol{y}_n\cdot\log{p(\hat{\boldsymbol{y}}_n|\boldsymbol{X}_n;\theta)} +\\ 
(1-\boldsymbol{y}_n)\cdot\log(1-p(\hat{\boldsymbol{y}}_n|\boldsymbol{X}_n;\theta)),
\end{aligned}
\end{equation}
where $\theta$ represents the parameters of the neural network, $\hat{\boldsymbol{y}}_n$ represents the estimated target.

Then, the total loss for training the student network is defined as a summary of the BCE-loss and the KD-loss weighted by hyperparameter $\alpha$:
\begin{equation}\label{eq7} 
\mathcal{L} = \mathcal{L}_{BCE} + \alpha \mathcal{L}_{KD}.
\end{equation}

During the initial training process, the probability estimates of multi-instances of student are handicapped, since the student hasn't built a robust mapping from the input to the output. The distillation loss may not be very useful or even be counterproductive \citep{Anil2018}. Thus, we adopt the gradually increased supervision from the teacher network by changing $\alpha$ in Eq. \ref{eq7},
\begin{equation}\label{eq8}
\alpha_{t} = \frac{1}{2}\alpha(1+\cos(\frac{T_{cur}}{T_{max}}\pi + \pi)),
\end{equation}
where $T_{cur}$ denotes the current number of epochs, and the ${T_{max}}$ denotes the max training epochs. \gh{The subscript \textit{t} means changing with time}. By doing this, the student network could gradually receive guidance from the teacher network. 

In addition to training the teacher and student sequentially, we investigate the Deep Mutual Learning (DML) \citep{Zhang2018} setting where the teacher and the student learn collaboratively and teach each other throughout the training process. Mutually matching the probability estimates can increase each network's posterior entropy, and help them achieve better generalization \citep{Zhang2018}. With deep mutual learning, multi-modal knowledge is expected to be thoroughly exploited by the student network.

\subsection{Implementation Details}\label{ms6}
In the student network, the visual feature maps (\gh{$\boldsymbol{I} \in \mathbb{R}^{18\times 22\times 18 \times 128}$}) were processed with global average pooling to produce 128-dimensional features. The teacher network generated the 128-dimensional features with attention weighted combining. The next two fully-connected (FC) layers had 128 and 16 units, respectively. The output layer had one unit for binary classification. Dropout with rate 0.4 was activated for the first FC layer. We used ELU \citep{Clevert2016} activation for the FC layers, and Sigmoid activation for the output layer to generate disease scores. We evenly divided each MRI image into 8 non-overlapping patches of size $72\times 88\times 72$. Note that the patch size of each axis was set as multiples of 8 to ensure that no border pixel was discarded during the convolution process. 

Our models were implemented with PyTorch (https://pytorch.org/). In addition, we defined our model in mixed-precision using the NVIDIA APEX library (https://github.com/NVIDIA/apex) to reduce the GPU memory. \gh{We used the batch size of 16 samples.} Each mini-batch consisted of balanced samples of each category. In order to obtain more reliable gradient estimates, we aggregated the gradients for a total of 4 mini-batches before performing a training step. We used the ADAM optimizer \citep{Kingma2015} to perform the update steps for all experiments. The learning rate was set as 2e-4. A constant learning rate was used since we did not found much improvement with learning rate decay. The training of an independent student or teacher took 30 or 20 epochs, respectively. The training of a student using KD-loss took 30 epochs with $\alpha=20$. In addition, the training under the DML setting took 30 epochs with $\alpha=150$.

Random flipping along the coronal axis and Gaussian noise are used for data augmentation. In addition, we use mixup \citep{Zhang2017} to generate convex combinations of pairs of examples and their labels for training the networks. Considering the small amounts of MCI subjects, we use AD and NC subjects as auxiliary data during the training process. Previous studies have shown the prediction of MCI conversion can be improved by the auxiliary knowledge learned from AD and NC \citep{Moradi2015,Lian2018}. We combine AD and pMCI as the positive group, since pMCI has similar atrophy patterns as AD. Meanwhile, the NC and sMCI subjects are combined as the negative group, as sMCI has less disease-related anomalous. With the help of mixup generating convex combinations of pairs of examples and their labels, the network is regularized with better generalizability to differentiate pMCI from sMCI. Note that the AD and NC subject are only used in the training process, both the validation and test set consist of only MCI subjects.

\section{Experiments and Analysis}\label{ea}
In this section, we first present the experimental settings and results. Next, we investigate the influences of the number of image patches, and the cumulative strategy of the KD-loss on prediction performance. Finally, we present the visual interpretation analysis.

\begin{table*}[!t]
\centering
\caption{Results of MCI conversion prediction on the ADNI-2 dataset.}\label{tab2}
\setlength{\tabcolsep}{3pt}
\begin{threeparttable}
\begin{tabular}{
p{40pt}<{\centering} | p{40pt}<{\centering}   p{40pt}<{\centering}   p{40pt}<{\centering}   p{40pt}<{\centering}   p{50pt}<{\centering} | p{40pt}<{\centering}   p{40pt}<{\centering}
}
\hline
\multirow{2}{*}{Method} & \multirow{2}{*}{Student} & 
\multirow{2}{*}{AT} & \multirow{2}{*}{SP} & 
M3ID & M3ID+DML & Clinical & \multirow{2}{*}{Teacher} \\ 
&&&& (Ours) & (Ours) & SVM & \\
\hline
&&&&&&&\\[-0.5em]
AUC (std.) & 0.822*\gh{†} (0.004)   & 0.838*\gh{†} (0.002) & 0.844*\gh{†} (0.002)       & {\ul 0.856* (0.006)} & \textbf{0.871 (0.004)} & 0.903 (0.002) & 0.912 (0.002) \\
&&&&&&&\\[-0.5em]
ACC (std.) & 0.752*\gh{†} (0.027)   & 0.764*\gh{†} (0.012) & 0.772*  (0.015)       & {\ul 0.780* (0.008)} & \textbf{0.800 (0.016)} & 0.821 (0.008) & 0.832 (0.017) \\
&&&&&&&\\[-0.5em]
SEN (std.) & 0.766   (0.031)   & 0.737*  (0.030) & {\ul 0.774   (0.025)} & 0.745* (0.034)  & \textbf{0.784 (0.007)} & 0.805 (0.052) & 0.829 (0.021) \\
&&&&&&&\\[-0.5em]
SPE (std.) & 0.743*\gh{†} (0.057)   & 0.783*  (0.014) & 0.770*\gh{†} (0.013)       & {\ul 0.804  (0.033)} & \textbf{0.810 (0.028)} & 0.831 (0.038) & 0.835 (0.017) \\
\hline
\end{tabular}
The best result is bolded and the second best is underlined. AT: attention transfer \citep{Zagoruyko2017}. SP: similarity preserving \citep{Tung2019}. Clinical SVM: a SVM classifier trained with clinical data. *: significantly different from the results with M3ID+DML (t-test, \textit{p} $<$ 0.05). \gh{†}: significantly different from the results with M3ID (t-test, \textit{p} $<$ 0.05).
\end{threeparttable}
\end{table*}

\subsection{Experimental Settings}
The proposed distillation approach was compared with similarity preserving (SP) \citep{Tung2019} and attention transfer (AT) \citep{Zagoruyko2017}. SP losses were computed using the features of the first FC layer, since we found the losses computed using the feature maps of the last convolution layer performed poorly. AT losses were computed using the feature maps of the ResNet and each of the two $A^{2}N$-blocks. We set the weight of the distillation loss for attention transfer as \gh{$\beta_{AT}=1000$}, and for similarity-preserving as \gh{$\beta_{SP}=100$}, respectively. Specifically, both SP and AT losses were computed with the features generated using whole MRI images as input. \gh{All the methods used the same auxiliary samples and the same data augmentation procedure.}

We used four metrics to evaluate the prediction performance: accuracy (ACC), sensitivity (SEN), specificity (SPE), and area under receiver operating characteristic curve (AUC). Before defining these metrics, we use TP, TN, FP, and FN to denote the true positive, true negative, false positive, and false negative values, respectively. Then we get ACC=(TP+TN)/(TP+TN+FP+FN), SEN=TP/(FN+TP), and SPE=TN/(TN+FP). The AUC is computed based on all possible pairs of SEN and 1-SPE, which are obtained by changing the thresholds performed on the prediction probabilities. AUC is a better metric than accuracy in imbalanced datasets and real-world applications, especially in clinical practice \citep{HajianTilaki2013,Huang2005,Bekkar2013}. 
\gh{We compared the results across different methods with two-sample t-test, using the one-tailed distribution.}

We first evaluated the models' prediction performance on the ADNI-2 dataset. The models were trained with MCI data in ADNI-1 and the auxiliary AD/NC data (Section \ref{res1}). Then, we exchanged the training and test dataset, and performed evaluations on the ADNI-1 dataset (Section \ref{res2}). Next, we conducted experiments to study the influences of the number of image patches (Section \ref{res3}), and the effectiveness of the cumulative strategy of the KD-loss (Section \ref{res4}). In the above experiments, 10\% of the training MCI samples were randomly selected for held-out validation. The model with the highest AUC value on the hold-out validation dataset was saved for evaluation.

\subsection{Results of MCI Conversion Prediction}\label{res1}
Table \ref{tab2} compares the prediction results obtained by the student network, teacher network, and the students trained with various knowledge distillation methods. The models were evaluated on MCI subjects of ADNI-2 dataset. The results evaluated on AIBL were shown in Table \ref{tabs6} in the \emph{Appendix}. We reported means and standard deviations of the results over the five model runs. The proposed M3ID method was used in two settings: 1) the teacher and student networks were trained sequentially (M3ID in Table \ref{tab2}); 2) the teacher and student networks were trained simultaneously with DML (M3ID+DML in Table \ref{tab2}). 

As shown in Table \ref{tab2}, the multi-modal teacher network achieved the highest prediction AUC, suggesting the effectiveness of complementary information provided by multi-modal data. We also trained a SVM classifier with the clinical data, which consisted of 13 scalars. The SVM classifier achieved an AUC of 0.903, which was much higher than the MRI-based student network. The high prediction performance using clinical data suggests the superior predictive value of clinical data over MRI data. On the other hand, this implies the demand of reducing the performance gap between using multi-modal data and single-modal MRI data, since the gathering of clinical data requires much human efforts and costs.

\begin{table*}[!t]
\centering
\caption{Results of MCI conversion prediction on the ADNI-1 dataset.}
\label{tab3}
\setlength{\tabcolsep}{3pt}
\begin{threeparttable}
\begin{tabular}{
p{40pt}<{\centering} | p{40pt}<{\centering}   p{40pt}<{\centering}   p{40pt}<{\centering}   p{40pt}<{\centering}   p{50pt}<{\centering} | p{40pt}<{\centering}   p{40pt}<{\centering}
}
\hline
\multirow{2}{*}{Method} & \multirow{2}{*}{Student} & 
\multirow{2}{*}{AT} & \multirow{2}{*}{SP} & 
M3ID & M3ID+DML & Clinical & \multirow{2}{*}{Teacher} \\ 
&&&& (Ours) & (Ours) & SVM & \\
\hline
&&&&&&&\\[-0.5em]
AUC (std.) & 0.777*\gh{†} (0.003)       & 0.788*\gh{†} (0.002)         & 0.794*\gh{†} (0.003)       & {\ul 0.805* (0.002)} & \textbf{0.808 (0.002)} & 0.821 (0.003) & 0.854 (0.008) \\
&&&&&&&\\[-0.5em]
ACC (std.) & 0.722*\gh{†} (0.008)       & \textbf{0.742  (0.015)} & 0.736   (0.014)       & 0.733  (0.007)       & {\ul 0.741 (0.020)}    & 0.753 (0.009) & 0.776 (0.022) \\
&&&&&&&\\[-0.5em]
SEN (std.) & 0.710 \gh{†} (0.007)       & \textbf{0.778  (0.054)} & {\ul 0.754  (0.021)} & 0.730  (0.022)       & 0.738 (0.034)          & 0.728 (0.050) & 0.760 (0.042) \\
&&&&&&&\\[-0.5em]
SPE (std.) & {\ul 0.744  (0.022)} & 0.682*\gh{†} (0.049)          & 0.706*\gh{†} (0.021)       & 0.738  (0.029)       & \textbf{0.746 (0.019)} & 0.786 (0.044) & 0.804 (0.047) \\
\hline
\end{tabular}
The best result is bolded and the second best is underlined. AT: attention transfer \citep{Zagoruyko2017}. SP: similarity preserving \citep{Tung2019}. Clinical SVM: a SVM classifier trained with clinical data. *: significantly different from the results with M3ID+DML (t-test, \textit{p} $<$ 0.05). \gh{†}: significantly different from the results with M3ID (t-test, \textit{p} $<$ 0.05).
\end{threeparttable}
\end{table*}

In Table \ref{tab2}, all knowledge distillation methods improved the MRI-based student network's prediction performance. Knowledge transferred from the multi-modal attention network helped the student network grasp more informative features. Compared with other distillation methods, our M3ID approach improved the student's prediction AUC by 1.8\% than using AT, by 1.2\% than using SP. This demonstrates the effectiveness of the proposed method for multi-modal knowledge transferring. When training the student in DML setting, the AUC was further improved to 0.871. This suggests that DML helps the student to better exploit the multi-modal knowledge during the training process, and generalize better to test data.

As we have discussed that the binary class outputs of a teacher network are not sufficiently informative to teach a student. We validated this by experiments. The results were presented in Table \ref{tabs3} and Table \ref{tabs4} of the \emph{Appendix}. Compared with our method, the vanilla KD using the whole image as a single patch achieved poorer results. We also conducted additional experiments using SP under the DML setting. The results in Table \ref{tabs8} show that our method achieved better results.

\subsection{Influence of Data Partition}\label{res2}
In the above experiments, we evaluated different methods on the baseline ADNI-2 dataset. To study the generalization of the proposed framework, we evaluated the models' prediction performance on baseline ADNI-1 dataset (trained with MCI in ADNI-2 and additional AD/NC). The prediction results were reported in Table \ref{tab3}. The results evaluated on AIBL were shown in Table \ref{tabs7} in the \emph{Appendix}. We reported means and standard deviations of the results over the five model runs.

The multi-modal teacher network again achieved the highest prediction AUC. The student trained with only data supervision (\textit{i.e.,} conversion labels) achieved an AUC of 0.777. With various multi-modal knowledge distillation methods, the student network's prediction AUC was increased by 1.1\% to 3.1\%, respectively. Our proposed distillation method again brought bigger AUC increases to the student network than the other methods. The student trained in DML setting achieved a higher prediction performance compared with the student trained with traditional setting of distillation. Although the AT and SP produced higher accuracy and sensitivity values, our methods achieved better AUC scores, which were of great importance in clinic and imbalanced datasets. The results in Table \ref{tab3} demonstrates the generalizability of the proposed framework. 

By comparing the results in Table \ref{tab2} and Table \ref{tab3}, it can be seen that the prediction performance on the ADNI-2 are better than the prediction performance on the ADNI-1. There might be two possible reasons: one is the insufficient training sample size, since there are more MCI samples in ADNI-1 than those in ADNI-2 (264 scans vs. 191 scans); the other is that the MCI subjects of these two datasets have different disease statuses (see Table \ref{tabs1} and Table \ref{tabs2} in the \emph{Appendix}), and the dataset itself could be the bottleneck of performance improvement. Meanwhile, our setting of using two independent dataset is closer to the real-world clinic situation, because a large existing dataset can be used for training models to be evaluated on newly enrolled patients.

\begin{figure}[t!]
\centerline{\includegraphics[width=1.0\columnwidth]{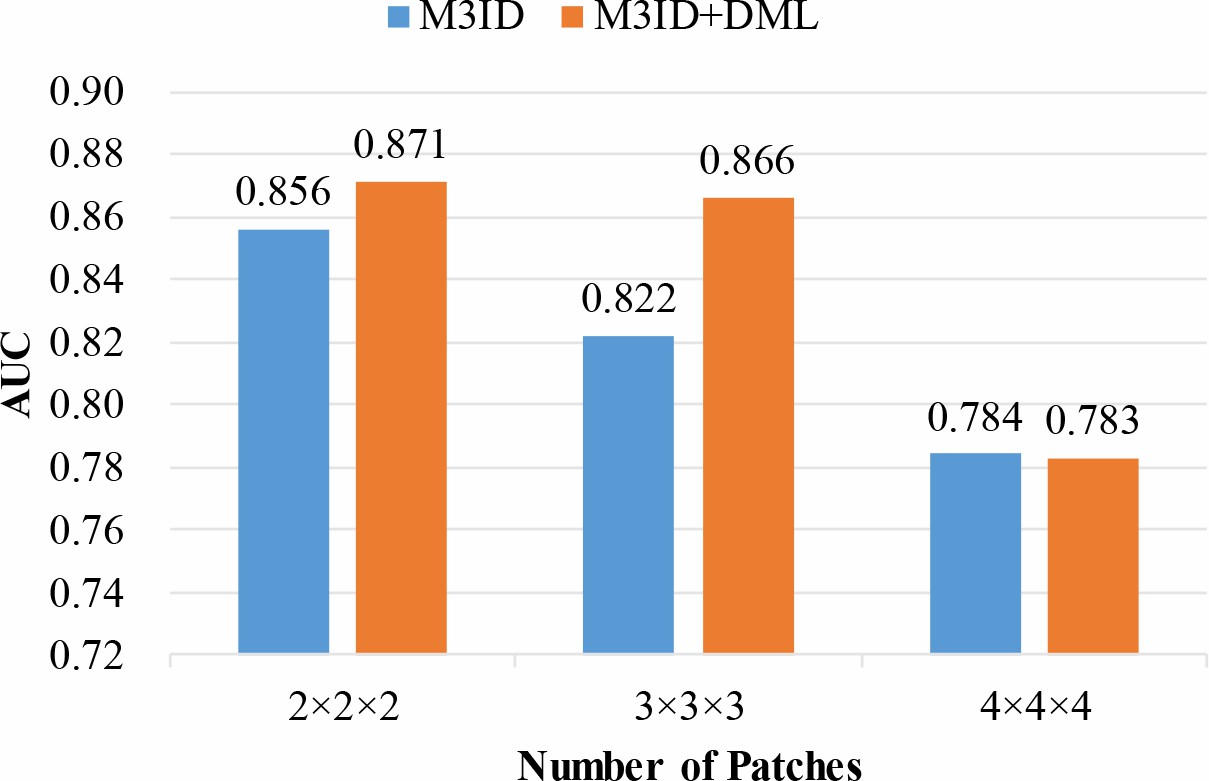}}
\setlength{\belowcaptionskip}{-10pt}
\caption{Influence of the number of image patches on the performance.}
\label{fig4}
\end{figure}

\subsection{Influence of The Number of Image Patches}\label{res3}
In the above experiments, we adopted a fixed number of patches (\textit{i.e.,} $2\times 2\times 2$) with size of $72\times 88\times 72$ for our proposed distillation method. We then explored the influence of the number of patches on the performance of knowledge distillation. Along each axis, we divided the whole image into 3 and 4 equal parts, producing 27 and 64 patches, respectively. Likewise, the patch size of each axis was multiples of 8 to ensure that no border pixel was discarded during the convolution process. When dividing an MRI into 27 patches, each patch is of size $48\times 64\times 48$. When dividing an MRI into 64 patches, each patch is of size $40\times 48\times 40$.

Fig. \ref{fig4} shows the AUC values of MCI conversion prediction on the ADNI-2 dataset. The best results were obtained by using 8 patches (\textit{i.e.,} $2\times 2\times 2$). When increasing the number of patches, the corresponding patch size decreases. Fig. \ref{fig4} suggests that the patches with much smaller sizes than the whole image are not suitable for transferring knowledge, since the structural information might be lost during the network's mapping process. In addition, using a large number of patches significantly increase the computational burden, which limits the proposed method's practical applications.

\subsection{Influence of the cumulative strategy of the KD-loss}\label{res4}
In this group of experiments, we investigated the influence of different cumulative strategies of the KD-loss on the performance. The prediction results on the ADNI-2 dataset were reported in Table \ref{tab4}: increasing the hyperparameter $\alpha$ from zero following the cosine function (Eq. \ref{eq8}) consistently brought performance improvements; the prediction performance with a constant $\alpha$ was comparatively poor. This suggests that the student network could learn better by gradually receiving more guidance from the teacher network. At the initial stage of training, a high knowledge distillation loss might hamper the accurate learning of a student \citep{Anil2018}. We conducted additional experiments with the linear cumulative strategy ($ \alpha_t=\alpha \times (1/T_{max})*T_{cur} $), and provided results in Table \ref{tab4} for comparison. The linear function also worked and produced comparable results. This suggests the effectiveness of our distillation method. The AUCs using cosine function (Eq. \ref{eq8}) are slightly higher than those using linear function. We conjecture this could result from the growth trend of the cosine function: starts with a smaller value when the network is handicapped; produces a higher value as the network getting robust.

To better understand the influence of $\alpha_t$, we provided change curves of the total loss, the KD-loss and the $\alpha_t$-weighted KD-loss during the training processes in Fig. \ref{figs1}. In addition, we presented the performance with different ${\alpha_t}$ in Fig. \ref{figs2} of the \emph{Appendix}.

\begin{table}[t!]
\centering
\caption{Influence of the cumulative strategy of the KD-loss on the performance.}\label{tab4}
\setlength{\tabcolsep}{3pt}
\begin{threeparttable}
\begin{tabular}{
p{30pt}<{\centering} | p{30pt}<{\centering} p{30pt}<{\centering} | p{30pt}<{\centering}   p{30pt}<{\centering} | p{30pt}<{\centering} p{30pt}<{\centering}
}
\hline
\multirow{3}{*}{Strategy} & \multicolumn{2}{c|}{Constant $\alpha$} 
& \multicolumn{2}{c|}{Linear $\alpha$} & \multicolumn{2}{c}{Cosine $\alpha$} \\
& M3ID & \begin{tabular}[c]{@{}c@{}} M3ID \\ +DML \end{tabular} 
& M3ID & \begin{tabular}[c]{@{}c@{}} M3ID \\ +DML \end{tabular}
& M3ID & \begin{tabular}[c]{@{}c@{}} M3ID \\ +DML \end{tabular} \\
\hline
&&&&\\[-1.0em]
AUC & 0.824 & 0.845          & 0.850 & {\ul 0.861}     
& 0.856 & \textbf{0.871} \\  &&&&\\[-1.0em]
ACC & 0.759 & 0.754          & 0.770 & \textbf{0.801}
& 0.791 & {\ul 0.796} \\     &&&&\\[-1.0em]
SEN & 0.697 & \textbf{0.803} & 0.776 & 0.776
& 0.711 & {\ul 0.789} \\     &&&&\\[-1.0em]
SPE & 0.800 & 0.722          & 0.765 & {\ul 0.817}
& \textbf{0.843} & 0.800 \\ 
\hline
\end{tabular}
The best result is bolded and the second best is underlined. 
\end{threeparttable}
\end{table}

\subsection{Visual interpretation analysis}\label{res5}
\begin{figure}[t!]
\centerline{\includegraphics[width=0.7\columnwidth]{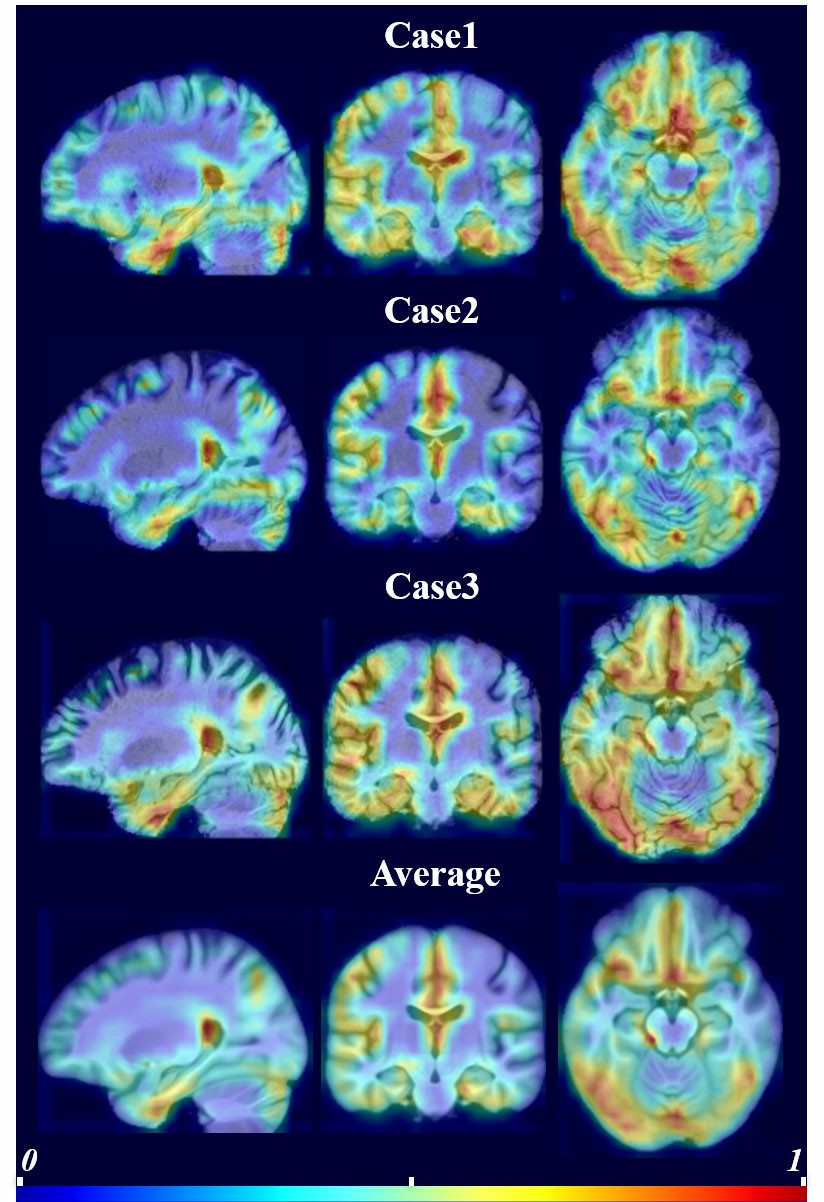}}
\caption{\ghr{Visualization of multi-modal attention maps. The attention maps are upscaled to the same size as the input MRI image, and overlayed onto the MRI image. The columns denote the views in three anatomical planes. The average results are computed using all correctly predicted test cases.}}
\label{fig5}
\end{figure}

In this group of experiments, we visualized the multi-modal attention maps and the patch scores for further analysis. Fig. \ref{fig5} shows the multi-modal attention maps generated by the teacher network. In Fig. \ref{fig5}, the hippocampus, ventricle, temporal and parietal lobe are highlighted. These areas have been previously reported in studies concerning the brain atrophy in AD \citep{Weiner2012}. This suggests that the multi-modal attention helps the teacher network to extract features from the discriminative areas.

\begin{figure*}[ht]
\centerline{\includegraphics[width=2.2\columnwidth]{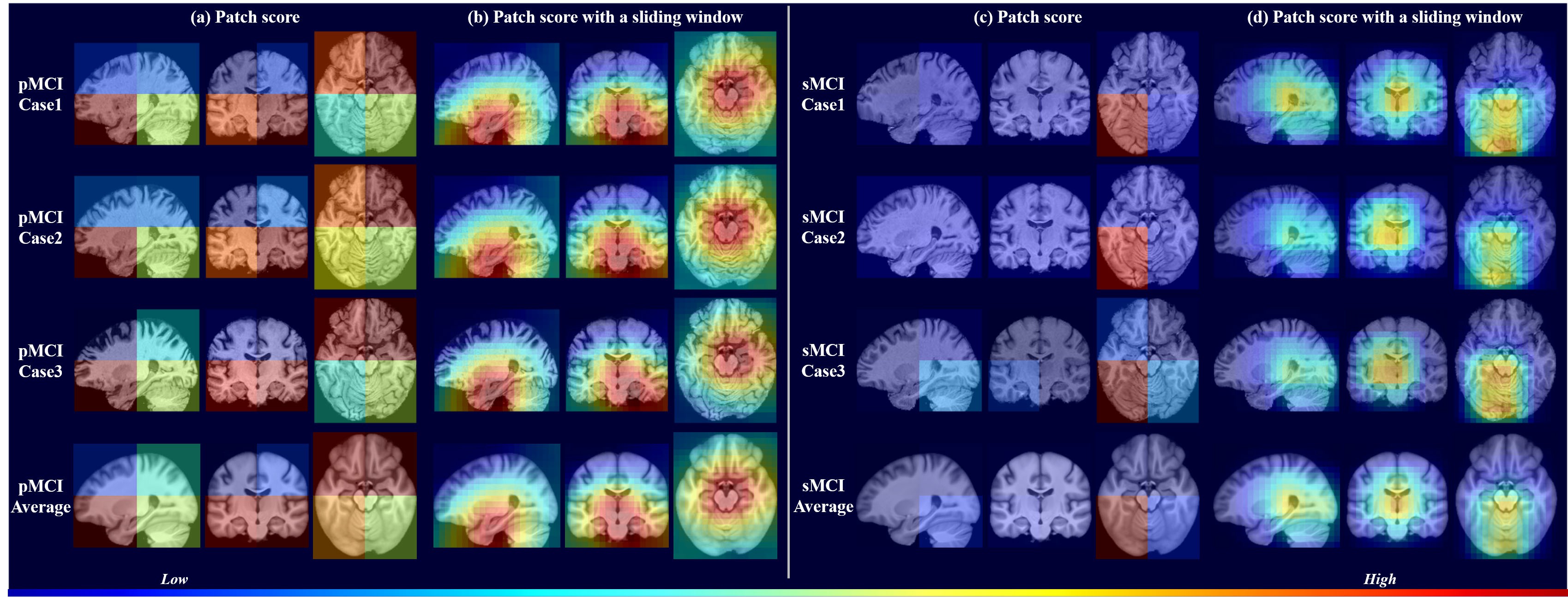}}
\caption{Visualization of patch scores of pMCI and sMCI subjects. Panel (a) and (c): 8 non-overlapping patches evenly divided from a whole MRI; panel (b) and (d): patches extracted with a sliding window running across the whole MRI. All of the patches are of size 72×88×72. The average results are computed using all correctly predicted test cases.}
\label{fig6}
\end{figure*}

\begin{figure*}[ht]
\centerline{\includegraphics[width=2.2\columnwidth]{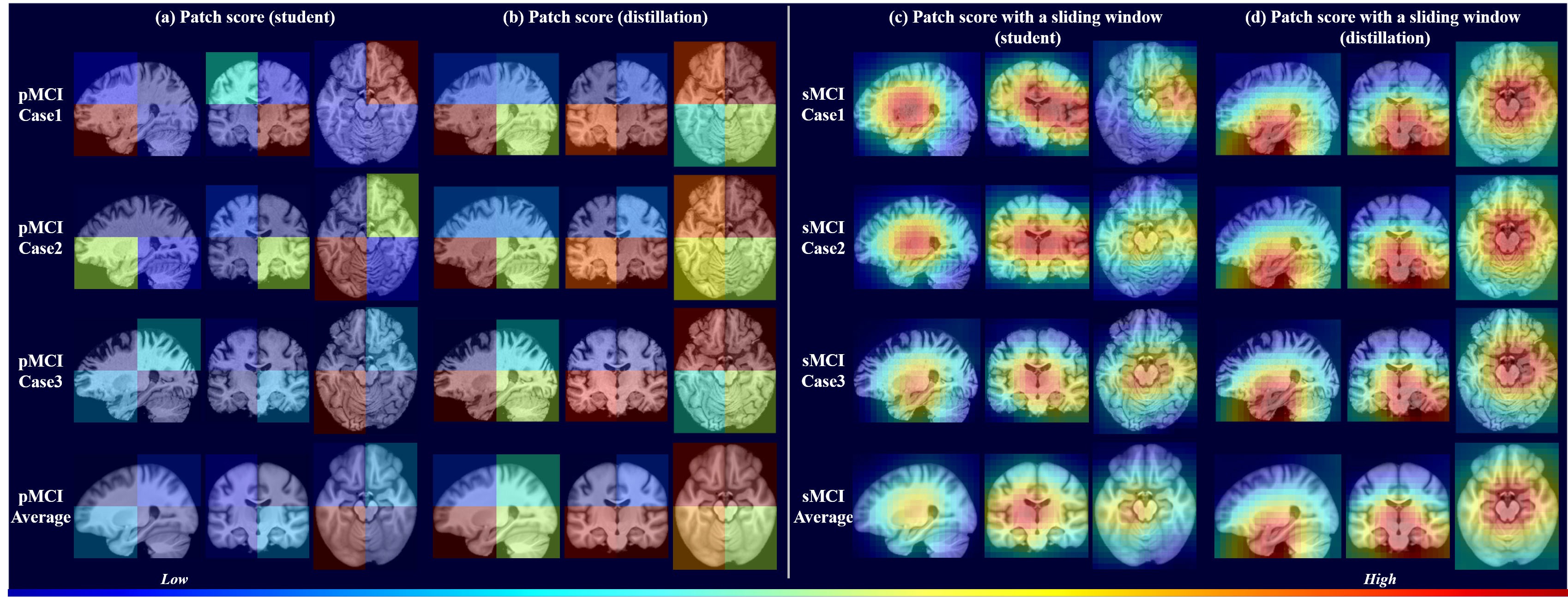}}
\caption{\gh{Visualization of patch scores generated without and with distillation. The student trained with distillation loss focus more on hippocampus and amygdala, while the vanilla student highlights diffuse areas across the brain. The 8 non-overlapping patches (of size 72×88×72) are evenly divided from a whole MRI. The average results are computed using all correctly predicted test cases.}}
\label{fig7}
\end{figure*}

The multi-instance probabilities are encoded to represent the complicated atrophy distributions of brains. With the multi-instance probabilities as extra guidance, the teacher network teaches the student network to reweight patches across the brain. Then the student has a better capacity of discriminative feature extraction and localization, which can be demonstrated in the visualizations of patch scores. Fig. \ref{fig6} visualizes the patch scores generated by the student network after distillation. For pMCI cases, the non-overlapping patches have different scores, and the patches near temporal lobe and hippocampus have higher scores. The disease-related areas like hippocampus and amygdala are located in the inner region of the temporal lobe. This demonstrates the model's capacity of discriminative feature extraction and localization. For sMCI subjects, patch scores are relatively low, and the highlighted areas are around the lateral ventricle. The visualization suggests that the network assesses regional severities of the diseases in MRI, and leverages the patch relationships for prediction. 

In addition, Fig. \ref{fig5} and Fig. \ref{fig6} show that the individual cases and the average results are largely overlapped. This suggests that the network can focus on the disease-related areas. By comparing Fig. \ref{fig5} and Fig. \ref{fig6}, we can see that the teacher's attention covers more areas than the student highlights. This could result from the multi-modal data providing complementary information. The teacher network uses an additional attention module to model the multi-modal correlations, for example, those between image and cognition. Thus the teacher network can captures more subtle patterns such as the structural changes of sulci (highlighted border areas of brains in Fig. \ref{fig5}), which are affected by AD and correlated to cognition \citep{Bertoux2019}. Although the guidance from teacher helps the student to focus more on the disease-related areas, some subtle patterns might be overly difficult for a student with less parameters and access to only MRI data. 

\gh{We compared the heat-maps generated by a student network trained without knowledge distillation (KD) and with KD in Fig. 7. The visualization results suggest that multi-instance knowledge distillation helps the network to reweight the importance of each patch, and focus on the disease-related areas. Then the student has a better capacity of discriminative feature extraction and localization, and achieves better performance (see Table 2 and Table 3).}
We presented additional figures in the \emph{Appendix} to further illustrate the effectiveness of our distillation method: Fig. \ref{figs3} shows that the teacher’s multi-instance probabilities reveals the patch relationships; 
Fig. \ref{figs6} visualizes the slices best showing the highlighted areas of sMCI. With our multi-instance distillation scheme, the network is capable of providing visual interpretations, which could be useful for medical decision makings.

\section{Discussion}\label{ds}
In this study, we present a novel knowledge distillation method for MCI conversion prediction. The multi-modal knowledge is transferred from a teacher network to a student network that only has access to MRI data. With the multi-modal knowledge as the extra supervision, the prediction performance of the MRI-based student network is effectively improved.

\subsection{Comparison with Previous Works}
In Table \ref{tab5}, we briefly summarized several state-of-the-art results on MCI conversion prediction using the ADNI dataset. We listed these method's results achieved with only MRI data, and the results using multi-modal data. The direct comparison may be affected because of the different number of subjects and the image pre-processing step used in these studies. Most previous studies (as shown in Table \ref{tab5}) used 36 months as the decision threshold between sMCI and pMCI. For a fair comparison, we also adopted 36 months to categorize sMCI and pMCI. Compared with the MRI-based prediction methods, our framework significantly increased the MRI-based prediction model's performance (see Table \ref{tab5}). Some previous studies' results \citep{Moradi2015,Spasov2019} using a combination of clinical data and MRI were much better than the performance achieved with only MRI data. Although this suggests the predictive values of clinical data, the disadvantages are obvious: the extra clinical data are necessary, and the highly accessible MRI data only play an auxiliary role. Our framework reduces the performance gap between using multi-modal data and using only MRI. Thus, our framework has promising potential application. Further, our experiments were conducted in a strict setting. The prediction model was trained and evaluated on three datasets (ADNI-1, ADNI-2, AIBL). With the more challenging evaluation protocol, the effectiveness of the proposed framework was ensured. 

\begin{table*}[!t]
\centering
\caption{A summary of MRI-based studies for predicting MCI-to-AD conversion using the ADNI dataset}\label{tab5}
\setlength{\tabcolsep}{3pt}
\begin{threeparttable}
\begin{tabular}{p{80pt}<{\centering} p{40pt}<{\centering} p{110pt}<{\centering} p{50pt}<{\centering} p{35pt}<{\centering} p{35pt}<{\centering} p{80pt}<{\centering} p{50pt}<{\centering}}
\hline

\multirow{2}{*}{Study} & \multirow{2}{*}{\begin{tabular}[c]{@{}c@{}} Number of \\ pMCI/sMCI \end{tabular}} & \multirow{2}{*}{Method} & \multirow{2}{*}{\begin{tabular}[c]{@{}c@{}} Conversion \\ time \end{tabular}} & \multicolumn{2}{c}{AUC (MRI)} & \multicolumn{2}{c}{\begin{tabular}[c]{@{}c@{}} AUC (MRI+) \end{tabular}} \\ &&&& ADNI-1 & ADNI-2 & ADNI-1 & ADNI-2 \\
\hline

\citet{Moradi2015} & 164/100 & \begin{tabular}[c]{@{}c@{}} Gray matter density map \\ + low density separation \end{tabular} & 36 months & 0.766 & - & \begin{tabular}[c]{@{}c@{}} 0.903 \\ (+ age \& clinical data) \end{tabular} & - \\

\citet{Liu2016}    & 117/117 & \begin{tabular}[c]{@{}c@{}} Gray matter density map \\ + Ensemble SVM \end{tabular} & 24 months & 0.834 & - & - & - \\

\citet{Liu2018}    & 205/465 & \begin{tabular}[c]{@{}c@{}} MRI patches \\ + CNN \end{tabular} & 36 months & - & 0.776 & - & - \\

\citet{Lian2018}   & 205/465 & \begin{tabular}[c]{@{}c@{}} MRI patches \\ + CNN \end{tabular} & 36 months & - & 0.781 & - & - \\

\citet{Lin2018}    & 169/139 & \begin{tabular}[c]{@{}c@{}} MRI patches + CNN \\ + PCA + Lasso + ELM \end{tabular} & 36 months & 0.829 & - & 0.861 (+ age) & - \\

\citet{Spasov2019} & 181/228 & \begin{tabular}[c]{@{}c@{}} Whole brain MRI \\ + CNN \end{tabular} & 36 months & 0.790 & - & \begin{tabular}[c]{@{}c@{}} 0.925 \\ (+ age \& clinical data) \end{tabular} & - \\

\citet{Pan2020}    & 192/502 & \begin{tabular}[c]{@{}c@{}} Whole brain MRI + CNN \\ with Fisher representation \end{tabular} & 36 months & - & 0.818 & - & 0.839 (+PET) \\
\hline

\multirow{2}{*}{Ours} & \multirow{2}{*}{240/215} & \multirow{2}{*}{Whole brain MRI + CNN} & \multirow{2}{*}{36 months} & \multirow{2}{*}{0.808} & \multirow{2}{*}{0.871} & 0.854 & 0.912 \\
&&&&&& \multicolumn{2}{c}{(+ age \& clinical data)}\\
\hline

\end{tabular}
``CNN'' refers to convolution neural network; ``MRI+'' refers to using multi-modal data in addition to MRI. ``ADNI-1'' and ``ADNI-2'' refer the corresponding evaluation datasets. 
\end{threeparttable}
\end{table*}

Many previous studies leverage the neurological expertise to define disease-related regions-of-interests (ROI), and extract features from these ROIs or nearby image patches for building diagnosis and prediction models \citep{Eskildsen2013,Moradi2015,Bron2015,Liu2018,Lian2018}. These studies demonstrate that the local image patches convey disease-related information, and the ROI priors are useful for better feature extraction. This is similar to our scheme that estimates probabilities of multi-instances. In our proposed scheme, the multi-instance probabilities are encoded to represent the complicated atrophy distributions. In conventional multi-instance learning \citep{Amores2013}, a bag is represented by the most discriminative instance or the average output of all the instances in the bag. By doing this, the distribution of atrophy patterns might be lost. In contrast, our method does not directly use multi-instance representations or outputs for classification, but considers the probability distribution of multi-instances as the extra supervision. With a better capacity of representing the complicated and distributed atrophy patterns, the multi-instance probabilities implicitly guide the student network to reweight the patches across a brain and achieve better performance. In addition, the multi-instance prediction models in \citep{Liu2018,Lian2018} require pre-defined landmarks for building an extra location proposal module, then perform feature extraction according to the proposed locations. In contrast, we use evenly partitioned image patches as input without the need of the localization processes or landmarks. Therefore, our deep learning framework is more annotation-efficient. 

Compared with the vanilla distillation method \citep{Hinton2015} that enforces the matching of class probabilities, our method enforce a matching of the probability estimates of multi-instances. The additional experiments demonstrate our method's advantage over the vanilla distillation method \citep{Hinton2015} (Table \ref{tabs3} and Table \ref{tabs4}) in the \emph{Appendix}. Compared with the representation-based distillation methods (including SP \citep{Tung2019} and AT \citep{Zagoruyko2017}) that encourage a student to mimic a teacher's representation space in different aspects, our method aims to match the probabilities of the output space. Considering the student network and the teacher network building with different inductive biases, their solution paths of expressing the representation space could be different. In addition, large gaps may exist between different sources of data, and lead to various responses in the intermediate layers. In the above situations, mimicking of representation space may not convey accurate knowledge. However, the differences in the representation space do not prevent the teacher and the student from independently making the right predictions. The experiments also demonstrate our method's superiority over the representation-based methods.

Our method lies between the probability-based method and representation-based distillation method: we use the multi-instance probabilities in the output space; the multi-instances probabilities work like a kind of representation rather than directly produce prediction outputs. 
\gh{In addition, our method could also be counted as multi-task learning: the student network is trained with the data supervision and the extra supervision distilled from the teacher network. Different from the conventional multi-task learning that uses clinical data as additional targets, we use the multi-modal teacher network’s predictions as the additional targets.}
We provide additional experimental analysis with respect to the effectiveness of temperature scaling and $A^{2}N$-block in the \emph{Appendix}. The results in Table \ref{tabs9} show that the temperature scaling brought no performance improvement. The prediction performance (see Table \ref{tabs10}) using the network without $A^{2}N$-block is poor, and our method still outperforms the other distillation methods.

\subsection{Limitations and Future Work}
Although the proposed framework has achieved competitive results for MRI-based conversion prediction, there is still room for improvement. Firstly, considering the medical data of high costs are used only during the training process, all available modalities of data can be leveraged to improve the current framework. For example, functional imaging, FDG-PET, and CSF data are useful for disease diagnosis and biomarker discovery \citep{Li2019,Wang2020,Wang2019,Ding2019}. We could further the study by using all available imaging and non-imaging clinical data for optimized knowledge extraction and distillation. Secondly, our teacher network only utilized the patients with complete multi-modal data. The patients missing some modalities of data were excluded. While in practice, modalities missing of medical data is usual due to the cost, data quality, and patient dropouts. Therefore, combining the methods of effectively utilizing incomplete multi-modal data into the proposed framework would be a promising direction. Thirdly, we processed the 3D patches without leveraging their location information, which could be another source of supervision. Explicitly using location information of these patches might further improve the model's ability for feature extraction. Finally, although we evaluated our frameworks on three datasets, further experiments in a larger population are needed to evaluate and improve the framework.

\section{Conclusion}\label{co}
In this study, we developed a multi-modal multi-instance distillation (M3ID) method for MRI-based MCI conversion prediction. We enforced an MRI-based student network to mimic a multi-modal teacher's probability estimates for 3D patches, and guide the student network to better model the subtle features in MRI. The proposed framework was evaluated on three public datasets. The experimental results shows that our M3ID approach effectively improved the prediction performance of the MRI-based student model. Our proposed framework enables better disease prediction using less medical examinations.

\section*{Acknowledgement}
Dr Chaoyue Wang is supported by the Australian Research Council Projects FL-170100117. We would like to thank the Alzheimer’s Disease Neuroimaging Initiative (ADNI) and the Australian Imaging, Biomarker \& Lifestyle Flagship Study of Ageing (AIBL) for data collection and sharing.


\bibliographystyle{model2-names.bst}
\biboptions{authoryear}
\bibliography{distill-nimg}

\beginappendix

\section*{Appendix}

\addcontentsline{toc}{section}{Appendix}
\renewcommand{\thesubsection}{\Alph{subsection}}

To ensure the effectiveness of the proposed multi-modal multi-instance distillation (M3ID) method for MCI conversion prediction, we conducted additional experiments with further analyses. We first showed the change curves of the knowledge distillation (KD) loss in Section \ref{s1}. 
In Section \ref{s2}, we presented the detailed clinical information of the studied subjects. 
In Section \ref{s3}, we showed the prediction performance using the whole MRI as a single patch. Then we presented the prediction results on the AIBL dataset in Section \ref{s4}. We also tested similarity preserving (SP) \citep{Tung2019} using the deep mutual learning (DML) \citep{Zhang2018} setting in Section \ref{s5}. Next in Section \ref{s6}, we investigated the influence of the temperature scaling on performance. In Section \ref{s7},  we evaluated predictive capability of the network without $A^{2}N$-blocks \citep{Chen2018}. In Section \ref{s8}, we showed how the network reveals the patch relationships. 
\gh{The effectiveness of the education variable was evaluated in Section \ref{s9}.
We investigated the effect of the scanner on the predicted results in Section \ref{s10}. The numbers of initially selected subjects regardless of MRI pre-processing were reported in Section \ref{s11}. We provided visualizations of sMCI patch scores in Section \ref{s12}. We showed the visualizations of incorrectly predicted cases in Section \ref{s13}.}

\subsection{Change curves of the losses during the training process}\label{s1}

\begin{figure}[ht!]
\centerline{\includegraphics[width=0.95\columnwidth]{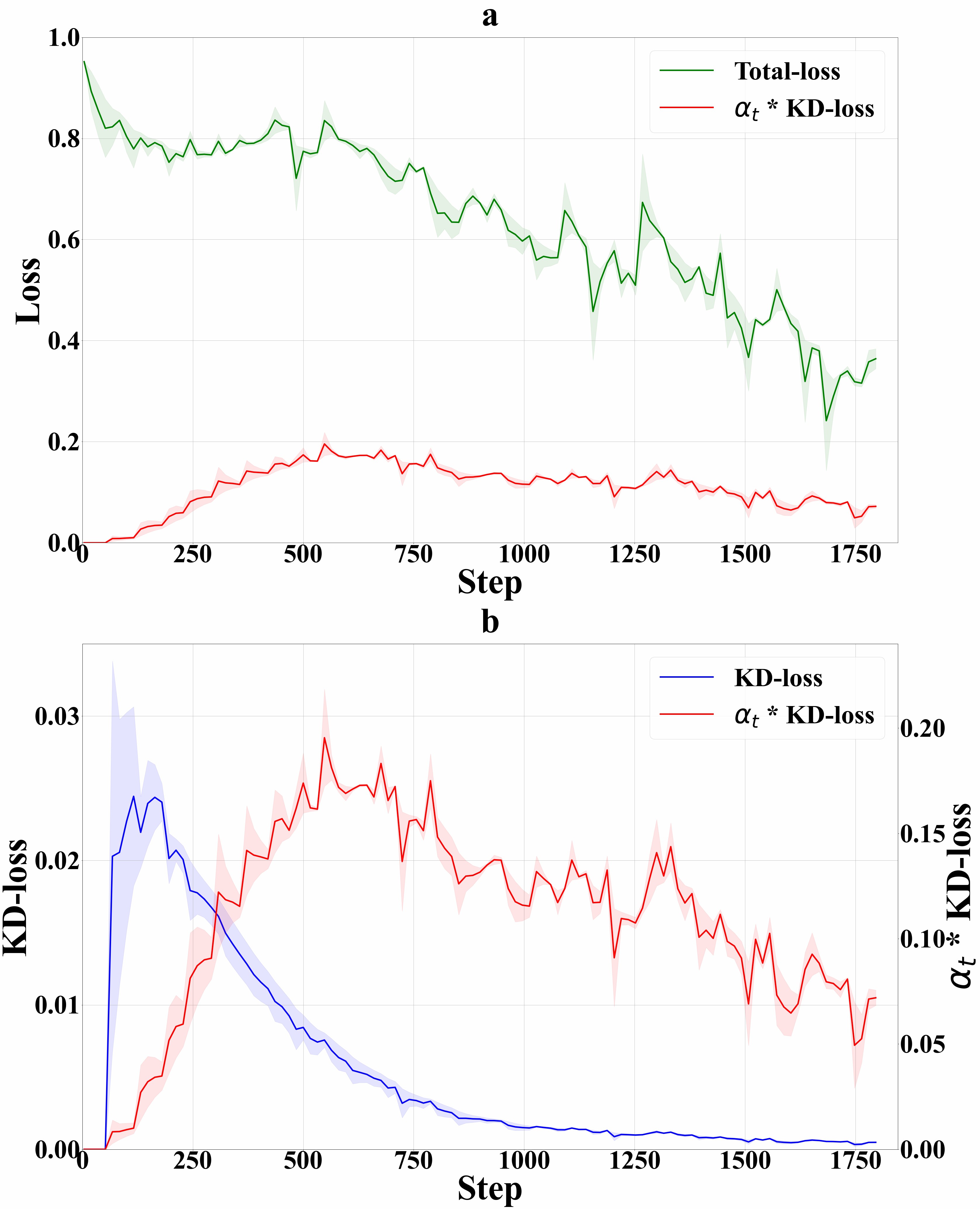}}
\setlength{\belowcaptionskip}{-10pt}
\caption{Panel a: Total-loss and $\alpha_t$-weighted KD-loss; panel b: KD-loss and $\alpha_t$-weighted KD-loss. The standard deviations of the losses are depicted in light color areas. The solid lines indicate the means of the losses.}
\label{figs1}
\end{figure}
\vspace{10pt}

\begin{table*}[t!]
\footnotesize
\centering
\caption{Detailed demographic and clinical information of the studied subjects.}
\label{tabs1}
\setlength{\tabcolsep}{3pt}
\begin{threeparttable}
\begin{tabular}{cccccccccccccccc}
\hline
\multirow{2}{*}{Category} & \multirow{2}{*}{Dataset} & \multirow{2}{*}{\begin{tabular}[c]{@{}c@{}}No. of\\ subjects\end{tabular}} & \multirow{2}{*}{\begin{tabular}[c]{@{}c@{}}Age\\ Range\end{tabular}} & \multirow{2}{*}{\begin{tabular}[c]{@{}c@{}}Females\\    /Males\end{tabular}} & \multirow{2}{*}{\begin{tabular}[c]{@{}c@{}}Years in\\   education\end{tabular}} & \multicolumn{3}{c}{\begin{tabular}[c]{@{}c@{}}APOe4\\   expression\end{tabular}} & \multirow{2}{*}{CDRSB} & \multirow{2}{*}{ADAS11} & \multirow{2}{*}{ADAS13} & \multicolumn{4}{c}{RAVLT}\\
&&&&&& 0 & 1 & 2 &&&& immediate & learning  & forgetting & \% forget\\
\hline
&&&&&& &&&&&&&\\[-0.5em]
\multirow{3}{*}{sMCI} & 
ADNI-1 & 100 & 57.8-87.9*  & 34/66   & 15.3 (3.2)  & 62  & 34 & 4  & 1.2 (0.6)  & 9.4 (3.8)   & 15.5 (5.5)  & 35.1 (9.7)   & 4.3 (2.6)  & 4.5 (2.4) & 55.3 (30.9) \\ & 
ADNI-2 & 115 & 55.0-88.6*  & 48/67   & 16.5 (2.7)† & 70  & 34 & 11 & 1.2 (0.7)  & 8.1 (3.5)†  & 12.8 (5.3)† & 39.4 (11.0)† & 5.2 (2.4)† & 4.3 (2.5) & 48.2 (30.7) \\ & 
AIBL   & 25  & 74.0-100.0* & 10/15   & - & - & -  & -  & - & - & - & - & - & - & - \\
&&&&&& &&&&&&&\\[-0.5em]
\multirow{3}{*}{pMCI} & 
ADNI-1 & 164 & 55.2-88.3*  & 66/98   & 15.6 (2.8)  & 53  & 83 & 28 & 1.9 (1.0)  & 13.2 (4.0)  & 21.4 (5.3)  & 27.0 (6.2)   & 2.7 (2.1)  & 4.8 (2.1) & 77.7 (28.1) \\ & 
ADNI-2 & 76  & 56.5-85.9*  & 36/40   & 16.4 (2.6)  & 20  & 41 & 15 & 2.2 (0.9)† & 14.1 (4.9)  & 22.5 (6.5)  & 28.4 (7.8)   & 3.2 (2.5)  & 5.3 (2.4) & 77.7 (27.8) \\ & 
AIBL   & 12  & 75.0-97.0*  & 4/8     & - & - & -  & -  & - & - & - & - & - & - & - \\
&&&&&& &&&&&&&\\[-0.5em]
\hline
&&&&&& &&&&&&&\\[-0.5em]
\multirow{2}{*}{NC}   
& ADNI-1 & 226 & 59.9-89.6   & 109/117 & 16.0 (2.9)  & 167 & 54 & 5  & 0.0 (0.1)  & 6.2 (2.9)   & 9.4 (4.2)   & 43.3 (9.1)   & 5.9 (2.3)  & 3.6 (2.7) & 34.1 (27.7) \\
& ADNI-2 & 186 & 56.2-89.0†  & 96/90   & 16.5 (2.5)  & 133 & 47 & 6  & 0.0 (0.1)  & 5.8 (3.1)   & 9.1 (4.5)   & 45.6 (10.5)† & 5.7 (2.4)  & 3.8 (2.6) & 36.9 (27.4) \\
&&&&&& &&&&&&&\\[-0.5em]
\multirow{2}{*}{AD}   
& ADNI-1 & 179 & 55.1-90.9   & 87/92   & 14.7 (3.2)  & 61  & 85 & 33 & 4.4 (1.6)  & 18.6 (6.3)  & 28.9 (7.6)  & 23.4 (7.4)   & 1.9 (1.7)  & 4.6 (1.8) & 88.6 (22.2) \\
& ADNI-2 & 140 & 55.6-90.3   & 56/84   & 15.8 (2.6)† & 45  & 66 & 29 & 4.5 (1.7)  & 20.7 (7.1)† & 31.0 (8.3)† & 22.5 (7.2)   & 1.9 (1.8)  & 4.5 (1.7) & 89.6 (19.3) \\
&&&&&& &&&&&&&\\[-0.5em]
\hline
\end{tabular}
The data is presented in a mean (std.) format. Abbreviations: APOe4,  Apolipoprotein E; CDRSB, Clinical Dementia Rating Sum of Boxes; ADAS, Alzheimer's disease Assessment Scale; RAVLT, Ray Auditory Verbal Learning Test.
* statistically significant difference in age between different datasets (ANOVA, $p < 0.05$).
† statistically significant difference between different datasets (t-test, $p < 0.05$).
‡ no statistically significant difference in sex between different datasets (Chi-Squared Test, $p > 0.05$).
\end{threeparttable}
\end{table*}
\vspace{-10pt}

Fig. \ref{figs1} shows the change curves of the Total-loss (a summary of $\alpha_t$-weighted KD-loss and BCE-loss) and the KD-loss (KL-divergence) during the training process. Fig. \ref{figs1} shows that as the training goes, the $\alpha_t$-weighted KD-loss increases at first then decrease, and the KD-loss keeps decreasing. This is attributed to the cumulative strategy of the KD-loss. During the training process, the hyper-parameter $alpha_t$ is increasing following the cosine function (Eq. \ref{eq8}). Yet the $\alpha_t$-weighted KD-loss and the Total-loss decrease in the later stage of training. This suggests that the student network could learn better by gradually receiving more guidance from the teacher network. 

\begin{figure}[ht!]
\centerline{\includegraphics[width=0.95\columnwidth]{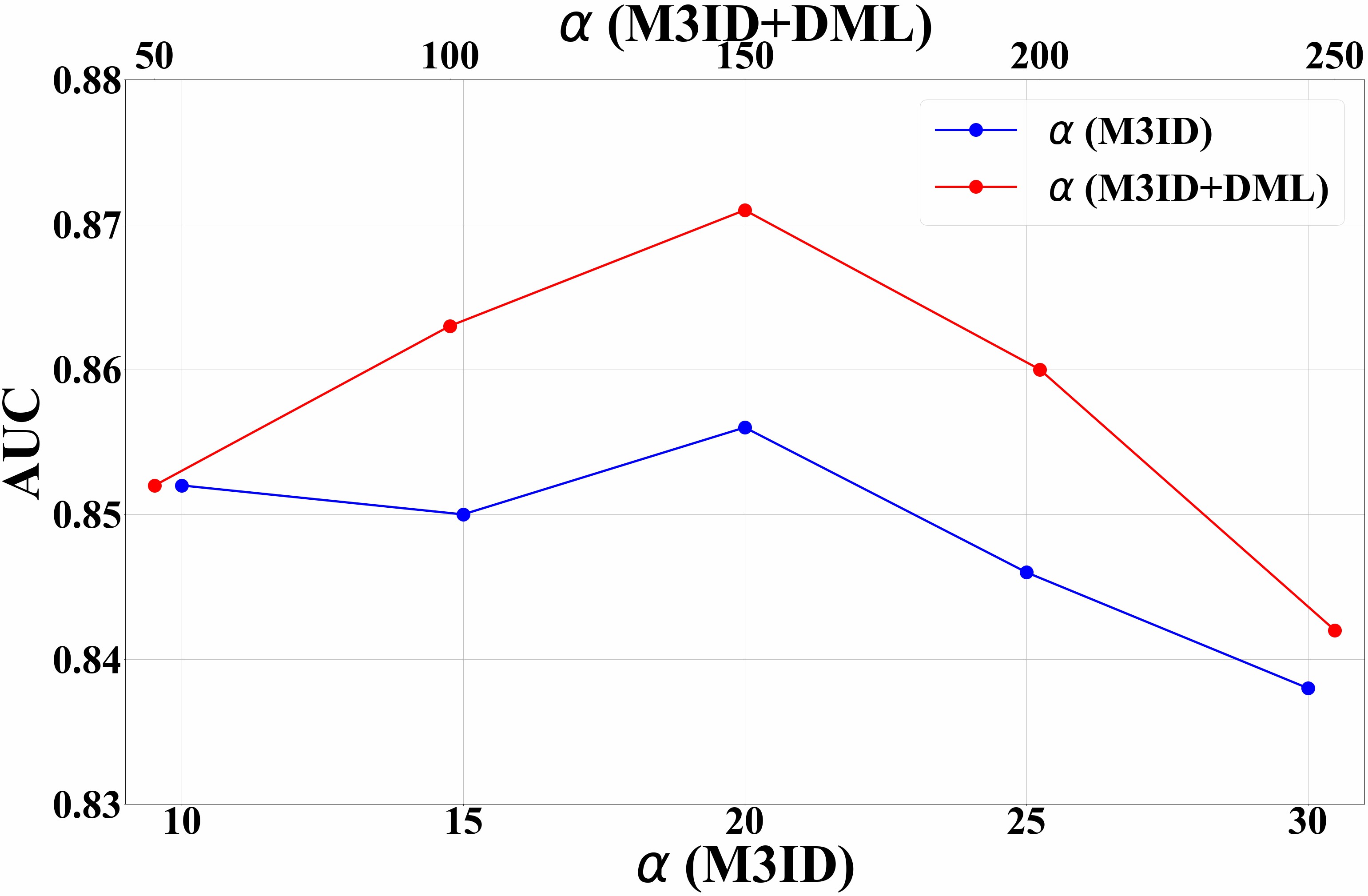}}
\setlength{\belowcaptionskip}{-10pt}
\caption{Influence of different $\alpha$ with the cumulative strategy on the performance.}
\label{figs2}
\end{figure}

In addition, we presented the performance with different $\alpha_t$ values (\textit{i.e.}, $\alpha$ with the cumulative strategy) in Fig. \ref{figs2}. Specifically, We set different $\alpha$ values and used the corresponding $\alpha_t$ values to weight the KD-loss (Eq. \ref{eq7}, \ref{eq8}). The performance curve with different $\alpha$ for M3ID is lined in blue. The performance curve with different $\alpha$ for M3ID+DML is lined in red.\\

\subsection{Detailed clinical information of the studied datasets}\label{s2}

Here we provide the detailed clinical information in Table \ref{tabs1}, which shows the differences between these two datasets in terms of demographic, neuropsychological and cognitive assessment as well as APOe4 genotyping data. The MCI subjects in ADNI-1 and ADNI-2 are different in many aspects. For example, the ADNI-1 subjects have a different gender ratio and a different conversion ratio. These differences implies that the MCI subjects of the two datasets have different disease statuses, which might restrict further performance improvements.

We conducted additional experiments using the SVM classifier on the balanced training and test sets. Specifically, we randomly under-sample both the training set and the test set for 10 times. After under-sampling, the training set and the test set have the same number of MCI subjects. We set the subject number of each class as 76 (the lowest subject number of pMCI is 76) for both the training set and the test set. The averaged results are given in Table \ref{tabs2}. The above table shows that the SVM classifier performed differently on ADNI-1 and ADNI2. This suggests that the effect of risk factors (\textit{i.e.}, the clinical indicators) varies in these two dataset, and the MCI subjects of these two datasets have different disease statuses.

\begin{table}[t!]
\footnotesize
\centering
\caption{Result of MCI conversion prediction using SVM classifier and clinical data on balanced training and test sets.}
\label{tabs2}
\setlength{\tabcolsep}{3pt}
\begin{threeparttable}
\begin{tabular}{p{35pt}<{\centering} p{35pt}<{\centering} | p{35pt}<{\centering} p{35pt}<{\centering} p{35pt}<{\centering} p{35pt}<{\centering}}
\hline
Train & Test & AUC (std) & ACC (std) & SEN (std) & SPE (std) \\
\hline
&&&&&\\[-0.5em]
\multirow{2}{*}{ADNI-1} & \multirow{2}{*}{ADNI-2} & 0.903 (0.007) & 0.805 (0.012) & 0.736 (0.019) & 0.874 (0.021) \\
&&&&&\\[-0.5em]
\multirow{2}{*}{ADNI-2} & \multirow{2}{*}{ADNI-1} & 0.822 (0.020) & 0.739 (0.019) & 0.787 (0.036) & 0.691 (0.015) \\
\hline
\end{tabular}
\end{threeparttable}
\end{table}

\subsection{Prediction performance using the whole MRI as a single patch}\label{s3}

We presented the experimental results with the whole MRI as input. Specifically, we used the vanilla KD \citep{Hinton2015} method. And the network's output layer was changed to have two output units, because the one dimensional output cannot be used to correctly calculate the KD-loss. The results on ADNI-1 and ADNI-2 datasets were presented in Table \ref{tabs3} and Table \ref{tabs4}, respectively. Compared with our method, the vanilla KD using the whole image as a single patch achieved poor results. This further demonstrates that the probabilities of two classes are not sufficiently informative to carry useful knowledge.

\begin{table}[t!]
\footnotesize
\centering
\caption{Results of MCI conversion prediction with vanilla KD and M3ID on the ADNI-2 dataset.}
\label{tabs3}
\setlength{\tabcolsep}{3pt}
\begin{threeparttable}
\begin{tabular}{c | c c c c }
\hline
Method & Vanilla KD & M3ID & \begin{tabular}[c]{@{}c@{}} Vanilla \\ KD+DML \end{tabular} & M3ID+DML \\
\hline
&&&&\\[-0.5em]
AUC (std.) & 0.840 (0.006) & 0.856 (0.006) & 0.844 (0.007)  & 0.871 (0.004) \\
&&&&\\[-0.5em]
ACC (std.) & 0.764 (0.012) & 0.780 (0.008) & 0.747 (0.023)  & 0.800 (0.016) \\
&&&&\\[-0.5em]
SEN (std.) & 0.760 (0.052) & 0.745 (0.034) & 0.792 (0.051)  & 0.784 (0.007) \\
&&&&\\[-0.5em]
SPE (std.) & 0.767 (0.035) & 0.804 (0.033) & 0.715 (0.066)  & 0.810 (0.028) \\
\hline
\end{tabular}
\end{threeparttable}
\end{table}

\begin{table}[t!]
\footnotesize
\centering
\caption{Results of MCI conversion prediction with vanilla KD and M3ID on the ADNI-1 dataset.}
\label{tabs4}
\setlength{\tabcolsep}{3pt}
\begin{threeparttable}
\begin{tabular}{c | c c c c }
\hline
Method  & Vanilla KD & M3ID & \begin{tabular}[c]{@{}c@{}} Vanilla \\ KD+DML \end{tabular} & M3ID+DML \\
\hline
&&&&\\[-0.5em]
AUC (std.) & 0.780 (0.006) & 0.805 (0.002) & 0.789 (0.003)  & 0.808 (0.002) \\
&&&&\\[-0.5em]
ACC (std.) & 0.713 (0.021) & 0.733 (0.007) & 0.721 (0.021)  & 0.741 (0.020) \\
&&&&\\[-0.5em]
SEN (std.) & 0.738 (0.033) & 0.730 (0.022) & 0.712 (0.039)  & 0.738 (0.034) \\
&&&&\\[-0.5em]
SPE (std.) & 0.672 (0.036) & 0.738 (0.029) & 0.736 (0.032)  & 0.746 (0.019) \\
\hline
\end{tabular}
\end{threeparttable}
\end{table}

\subsection{Prediction performance on the AIBL dataset}\label{s4}
\vspace{10pt}


\begin{table}[ht!]
\scriptsize
\centering
\caption{Results of MCI conversion prediction on the AIBL dataset. The models were trained with MCI data in ADNI-1 and the auxiliary AD/NC data.}
\label{tabs6}
\setlength{\tabcolsep}{3pt}
\begin{threeparttable}
\begin{tabular}{c | c c c c c}
\hline
Method     & Student       & AT            & SP            & M3ID          & M3ID+DML        \\
\hline
&&&&&\\[-0.5em]
AUC (std.) & 0.742 (0.005) & 0.725 (0.005) & 0.751 (0.003) & 0.763 (0.008) & 0.795   (0.004) \\
&&&&&\\[-0.5em]
ACC (std.) & 0.692 (0.015) & 0.703 (0.043) & 0.703 (0.033) & 0.719 (0.024) & 0.725   (0.035) \\
&&&&&\\[-0.5em]
SEN (std.) & 0.650 (0.038) & 0.667 (0.059) & 0.684 (0.037) & 0.700 (0.045) & 0.700   (0.045) \\
&&&&&\\[-0.5em]
SPE (std.) & 0.712 (0.033) & 0.720 (0.040) & 0.712 (0.044) & 0.728 (0.052) & 0.736   (0.036) \\
\hline
\end{tabular}
\end{threeparttable}
\end{table}
\vspace{-5pt}

\begin{table}[ht!]
\scriptsize
\centering
\caption{Results of MCI conversion prediction on the AIBL dataset. The models were trained with MCI data in ADNI-2 and the auxiliary AD/NC data.}
\label{tabs7}
\setlength{\tabcolsep}{3pt}
\begin{threeparttable}
\begin{tabular}{c | c c c c c}
\hline
Method     & Student       & AT            & SP            & M3ID          & M3ID+DML      \\
\hline
&&&&&\\[-0.5em]
AUC (std.) & 0.739 (0.003) & 0.749 (0.004) & 0.759 (0.002) & 0.779 (0.008) & 0.800 (0.007) \\
&&&&&\\[-0.5em]
ACC (std.) & 0.687 (0.036) & 0.687 (0.031) & 0.692 (0.041) & 0.714 (0.024) & 0.735 (0.023) \\
&&&&&\\[-0.5em]
SEN (std.) & 0.650 (0.038) & 0.650 (0.070) & 0.700 (0.045) & 0.750 (0.083) & 0.767 (0.037) \\
&&&&&\\[-0.5em]
SPE (std.) & 0.704 (0.067) & 0.704 (0.022) & 0.688 (0.066) & 0.696 (0.046) & 0.720 (0.028) \\
\hline
\end{tabular}
\end{threeparttable}
\end{table}

We conducted additional evaluations on the Australian Imaging, Biomarkers and Lifestyle (AIBL\footnote{The AIBL researchers contributed data but did not participate in analysis or writing of this report. AIBL researchers are listed at \url{www.aibl.csiro.au.}}) dataset \citep{Ellis2009}, which has a similar study goal and MRI scanning protocols to those used for ADNI. Most of the ADNI MCI subjects were followed every 6 months, while the AIBL MCI subjects were followed every 18 months. We screened out 37 AIBL MCI subjects for evaluation. The demographic characters of these MCI subjects were presented in Table \ref{tabs1}. Considering AIBL did not provide all the clinical measures as the ADNI did, we did not evaluate the multi-modal teacher model and the clinical SVM classifier on the AIBL dataset. We compared our methods with attention transfer (AT) \citep{Zagoruyko2017}, similarity preserving (SP) \citep{Tung2019}. In addition, we tested the effectiveness of our method using the deep mutual learning (DML) \citep{Zhang2018} setting. The results were summarized in Table \ref{tabs6} and Table \ref{tabs7}. On the AIBL dataset, our method achieved the best prediction performance. The experiments further demonstrate our method's generalizability and superiority over other methods.

\subsection{Prediction performance using SP under the DML setting}\label{s5}

\begin{table}[ht!]
\footnotesize
\centering
\caption{Comparisons of different KD methods using the DML setting. The models were evaluated on the ADNI-2 dataset.}
\label{tabs8}
\setlength{\tabcolsep}{3pt}
\begin{threeparttable}
\begin{tabular}{c | cccc}
\hline
Method     & SP            & SP+DML        & M3ID          & M3ID+DML      \\
\hline
&&&\\[-0.5em]
AUC (std.) & 0.844 (0.002) & 0.833 (0.006) & 0.856 (0.006) & 0.871 (0.004) \\
&&&\\[-0.5em]
ACC (std.) & 0.772 (0.015) & 0.765 (0.012) & 0.780 (0.008) & 0.800 (0.016) \\
&&&\\[-0.5em]
SEN (std.) & 0.774 (0.025) & 0.718 (0.033) & 0.745 (0.034) & 0.784 (0.007) \\
&&&\\[-0.5em]
SPE (std.) & 0.770 (0.013) & 0.797 (0.016) & 0.804 (0.033) & 0.810 (0.028) \\
\hline
\end{tabular}
\end{threeparttable}
\end{table}

According to a recent study \citep{Tian2020}, the representation-based methods (including SP and AT) perform worse when using the DML setting for training the student, while the probability-based methods (for example, the vanilla KD \citep{Hinton2015}) perform better. In \citep{Tian2020}, the authors conjecture that during the mutual-learning process, the probability-based methods do not require the output probabilities to be very accurate, since these methods perform like an advance label smoothing regularization. But the representation-based distillation methods are hard to learn knowledge from the handicapped teacher model. We conducted experiments using SP under the DML setting. The results in Table \ref{tabs8} also show that SP performed worse when using the DML setting for training the student.

\subsection{Influence of the temperature scaling on the performance}\label{s6}

\begin{table}[ht!]
\footnotesize
\centering
\caption{Influence of the temperature scaling on the performance. The models were evaluated on the ADNI-2 dataset.}
\label{tabs9}
\setlength{\tabcolsep}{3pt}
\begin{threeparttable}
\begin{tabular}{c | c c c | c c c}
\hline
\multirow{2}{*}{\begin{tabular}[c]{@{}c@{}}Temperature \\ Scaling\end{tabular}} 
& \multicolumn{3}{c}{M3ID} & \multicolumn{3}{c}{M3ID+DML} \\ 
& T=1 & T=2 & T=4 & T=1 & T=2 & T=4 \\
\hline
&&&&&&\\[-0.5em]
AUC & 0.856  & 0.840  & 0.827  & 0.871    & 0.855   & 0.848  \\
&&&&&&\\[-0.5em]
ACC & 0.791  & 0.759  & 0.749  & 0.796    & 0.791   & 0.770  \\
&&&&&&\\[-0.5em]
SEN & 0.711  & 0.671  & 0.697  & 0.789    & 0.776   & 0.789  \\
&&&&&&\\[-0.5em]
SPE & 0.843  & 0.817  & 0.783  & 0.800    & 0.800   & 0.757  \\
\hline
\end{tabular}
\end{threeparttable}
\end{table}
\vspace{-10pt}

According to the vanilla KD paper \citep{Hinton2015}, the relative probabilities of incorrect classes - the class correlations - convey useful information about how a teacher tends to generalize. However, the probabilities of incorrect classes have very little influence on the cost function because they are so close to zero. The temperature scaling is used for producing ``soft targets", which means a softer probability distribution over classes. Thus the knowledge transfer can be more effective. 

\begin{figure*}[t!]
\centerline{\includegraphics[width=1.5\columnwidth]{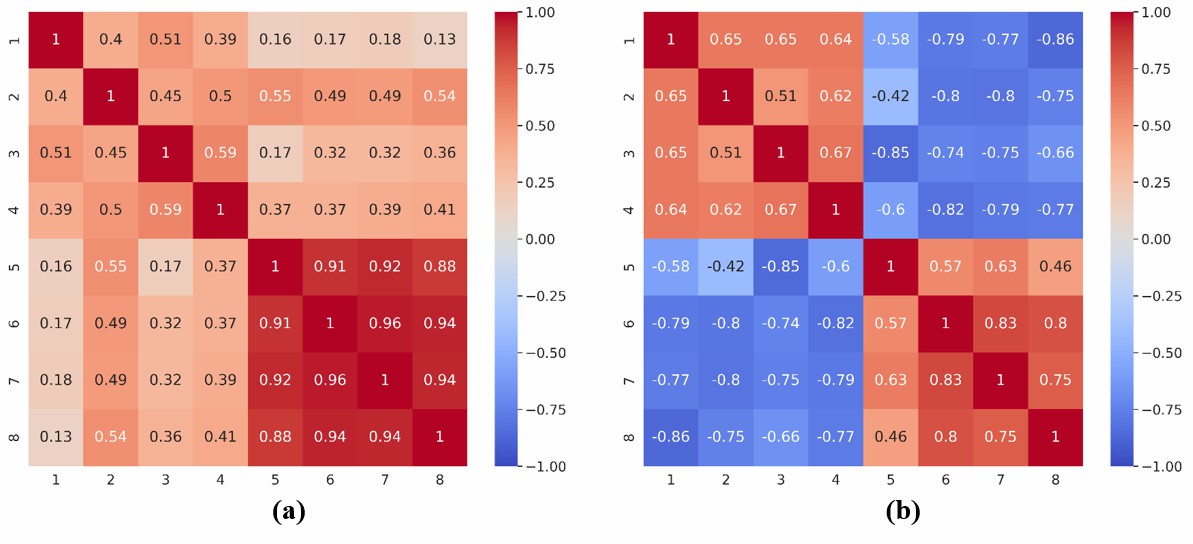}}
\setlength{\belowcaptionskip}{-10pt}
\caption{Pearson correlations of teacher network’s output multi-instance probabilities: (a) raw scores, (b) after-softmax probabilities. Every 4 patches are within the same part of brain: 1, 2, 3, 4 are of the lower half-part; 5, 6, 7, 8 are of the upper half-part. The disease-related areas such as amygdala and hippocampus lay in the lower half part of the brain. Patch-correlations can be interpreted as the color ``square''.}
\label{figs3}
\end{figure*}

\begin{figure}[h!]
\centerline{\includegraphics[width=0.75\columnwidth]{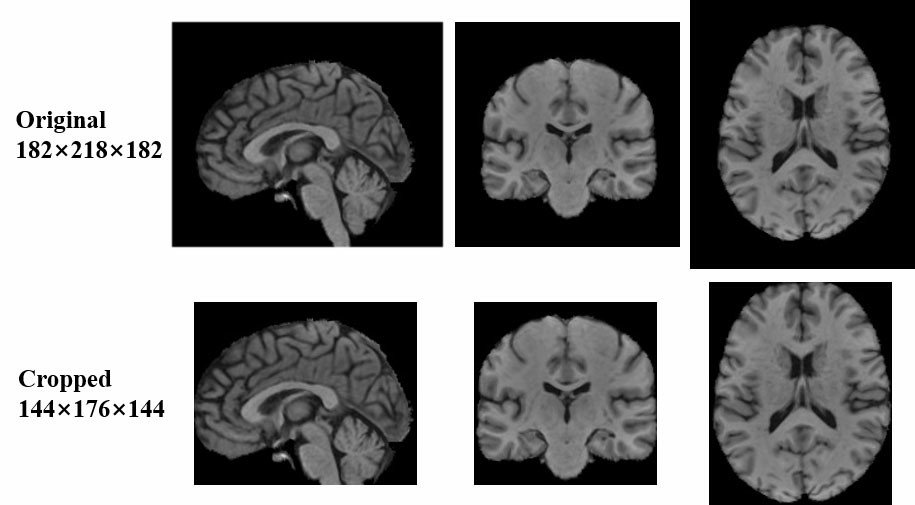}}
\setlength{\belowcaptionskip}{-10pt}
\caption{MRI in original size and cropped size. We choose the cropped size (144×176×144) to cover most areas of the brain, and reduce the useless computation in the background.}
\label{figs7}
\end{figure}

While in our method, the multi-instance outputs are used as a kind of intermediate representation rather than to produce predictions. The bags of multi-instances do not have corresponding classes since we didn't define positive or negative bags (like the manner used in multi-instance learning). In addition, if we consider the probability of each instance as the estimate of a certain class, our model's probability distribution could be soft enough. This is shown in the Fig. \ref{fig6} and Fig. \ref{figs3}, in which several patches have scores higher than zero and are visualized with warm colours. Therefore, we believe the hyper-parameter is inessential to our method. 

The authors of \citep{Hinton2015} found that increasing the temperature did not always increase the model's performance. We also tested the effectiveness of the temperature scaling. The results in Table \ref{tabs9} show that the temperature scaling brought no performance improvement. In addition, a method would be unpractical if there are too many hyper-parameters to be carefully tuned. So we didn't use this hyper-parameter.

\subsection{Prediction performance using the network without $A^{2}N$-blocks}\label{s7}

\begin{table}[ht!]
\footnotesize
\centering
\caption{Results of MCI conversion prediction with and without $A^{2}N$-blocks. The models were evaluated on the ADNI-2 dataset.}
\label{tabs10}
\setlength{\tabcolsep}{3pt}
\begin{threeparttable}
\begin{tabular}{c | c c c c c | c}
\hline
Method & Student & AT    & SP    & M3ID  & \begin{tabular}[c]{@{}c@{}} M3ID \\ +DML \end{tabular} & Teacher \\
\hline
&&&&&&\\[-0.5em]
\begin{tabular}[c]{@{}c@{}} AUC without \\ $A^{2}N$-blocks \end{tabular} & 0.739 & 0.733 & 0.743 & 0.753 & 0.767 & 0.904   \\
&&&&&&\\[-0.5em]
\begin{tabular}[c]{@{}c@{}} AUC with    \\ $A^{2}N$-blocks \end{tabular} & 0.822 & 0.839 & 0.844 & 0.856 & 0.871 & 0.915   \\
\hline
\end{tabular}
\end{threeparttable}
\end{table}

We conducted additional experiments to show the efficacy of $A^{2}N$-block \citep{Chen2018}. The prediction performance was evaluated on the ADNI-2 dataset, and the results were given in Table \ref{tabs10}. Results in Table \ref{tabs10} show that the network without $A^{2}N$-block performed poorly. This demonstrates the effectiveness of $A^{2}N$-block for extracting better features from MRI. Considering the limited sample size and the high dimension of MRI data, we chose to use a light-weight network for reducing the risk of overfitting. We devised the network with $A^{2}N$-blocks cascading on a shallow 3D-ResNet. In addition, our method still outperforms other distillation methods Table \ref{tabs10}. This further demonstrates the effectiveness of our method.

\begin{table}[b!]
\footnotesize
\centering
\caption{Results of MCI conversion prediction with and without the Education variable.}
\label{tabs11}
\setlength{\tabcolsep}{3pt}
\begin{threeparttable}
\begin{tabular}{c | c c}
\hline
Method                & SVM           & CNN           \\
\hline
&&\\[-0.5em]
AUC without education & 0.901 (0.002) & 0.822 (0.004) \\
&&\\[-0.5em]
AUC with education    & 0.903 (0.002) & 0.852*(0.008) \\
\hline
\end{tabular}
* significantly different from the results with education-involved method (t-test, $p < 0.05$).
\end{threeparttable}
\end{table}

\begin{table*}[t!]
\centering
\caption{Numbers of MRI scanned by different models of scanners. "m0-19" refer to 20 different models of scanners.}
\label{tabs12}
\setlength{\tabcolsep}{3pt}
\begin{threeparttable}
\begin{tabular}{cccccccccccccccccccccc | c}
\hline
\multirow{2}{*}{Category} & \multirow{2}{*}{Dataset} & \multicolumn{20}{c}{Scanner} & \multirow{2}{*}{Total} \\
&& m0 & m1 & m2 & m3 & m4 & m5 & m6 & m7 & m8 & m9 & m10 & m11 & m12 & m13 & m14 & m15 & m16 & m17 & m18 & m19 &                              \\
\hline
\multirow{4}{*}{ADNI-1}   
& sMCI & 2  & 6  & 0  & 0  & 0  & 12 & 0  & 0  & 0  & 6  & 1   & 39  & 7   & 0   & 0   & 9   & 3   & 15  & 0   & 0   & 100                    \\
& pMCI & 2  & 14 & 0  & 0  & 0  & 15 & 3  & 0  & 0  & 13 & 0   & 58  & 2   & 0   & 0   & 23  & 1   & 33  & 0   & 0   & 164                    \\
& NC   & 3  & 19 & 0  & 0  & 0  & 19 & 5  & 0  & 0  & 19 & 2   & 87  & 2   & 0   & 0   & 24  & 1   & 45  & 0   & 0   & 226                    \\
& AD   & 1  & 11 & 0  & 0  & 0  & 15 & 4  & 0  & 0  & 15 & 3   & 69  & 8   & 0   & 0   & 21  & 1   & 31  & 0   & 0   & 179                    \\
\multirow{4}{*}{ADNI-2}   
& sMCI & 15 & 0  & 8  & 3  & 1  & 0  & 0  & 1  & 0  & 5  & 0   & 0   & 3   & 11  & 4   & 0   & 0   & 0   & 35  & 29  & 115                    \\
& pMCI & 9  & 0  & 14 & 2  & 1  & 0  & 0  & 3  & 0  & 3  & 0   & 0   & 0   & 6   & 7   & 0   & 0   & 0   & 22  & 9   & 76                     \\
& NC   & 22 & 0  & 15 & 0  & 3  & 0  & 0  & 4  & 0  & 6  & 0   & 0   & 4   & 31  & 13  & 0   & 0   & 0   & 73  & 15  & 186                    \\
& AD   & 22 & 0  & 19 & 3  & 0  & 0  & 0  & 5  & 1  & 3  & 0   & 0   & 0   & 22  & 12  & 0   & 0   & 0   & 43  & 10  & 140                    \\
\hline
\multicolumn{2}{c}{Total}
& 76 & 50 & 56 & 8  & 5  & 61 & 12 & 13 & 1  & 70 & 6   & 253 & 26  & 70  & 36  & 77  & 6   & 124 & 173 & 63  & 1186                         \\
\hline
\end{tabular}
\end{threeparttable}
\end{table*}
\vspace{-10pt}

\subsection{Patch relationships}\label{s8}

The patch relationships, which reflect the atrophy distribution, are critical for knowledge transfer. Using all correctly predicted test cases, we plot Pearson correlations of output probabilities before and after softmax. Fig. \ref{figs3}-(b) shows that the teacher’s multi-instance probabilities (after-softmax) reveals the patch relationships: disease-related patches (patch 1-4) have positive correlations to each other; the disease-related patches (patch 1-4) and the other patches (patch 5-8) are negatively correlated. Hence distillation encourages the student to capture such relationships. While Fig. \ref{figs3}-(a) shows the raw scores correlations: the correlations between the disease-related patches (patch 1-4) are weaker in comparison to the correlations in Fig. \ref{figs3}-(b); and the disease-related patches (patch 1-4) and the other patches (patch 5-8) are moderately correlated. This observation verifies that the after-softmax probabilities help reveal patch relationships.

\subsection{The effectiveness of the education variable}\label{s9}

Previous studies found the positive correlation between education and brain volume in MCI subjects \citep{Wada2018}. We performed additional experiments to examine the effectiveness of education in predicting MCI conversion. First, we trained a SVM without education (using 12 variables). Then we trained a convolutional neural network (CNN) with MRI and only education as the clinical feature. The CNN architecture is the same as the teacher network described in the manuscript. The results are reported in Table \ref{tabs11}. The SVM classifier produces similar results whether or not using education. In contrast, the CNN with education produces better results compared to the CNN without education. We find no statistical difference in education between difference MCI groups (t-test, $p > 0.05$). Thus, education might be a useful AD indicator, which is correlated with MRI.

\begin{table}[ht!]
\centering
\caption{Top-6 scanners and corresponding images in ADNI-1.}
\label{tabs13}
\setlength{\tabcolsep}{3pt}
\begin{threeparttable}
\begin{tabular}{cccccccc | c}
\hline
\multirow{2}{*}{Category} & \multirow{2}{*}{Dataset} & \multicolumn{6}{c}{Scanner}    & \multirow{2}{*}{Total} \\
&& m1 & m5 & m9 & m11 & m15 & m17 &                        \\ 
\hline
\multirow{4}{*}{ADNI-1}   & sMCI                     & 6  & 12 & 6  & 39  & 9   & 15  & 87                     \\
                          & pMCI                     & 14 & 15 & 13 & 58  & 23  & 33  & 156                    \\
                          & NC                       & 19 & 19 & 19 & 87  & 24  & 45  & 213                    \\
                          & AD                       & 11 & 15 & 15 & 69  & 21  & 31  & 162                    \\
\hline
\multicolumn{2}{c}{Total}                            & 50 & 61 & 53 & 253 & 77  & 124 & 618 \\
\hline
\end{tabular}
\end{threeparttable}
\end{table}

\begin{table}[ht!]
\centering
\caption{Top-6 scanners and corresponding images in ADNI-2.}
\label{tabs14}
\setlength{\tabcolsep}{3pt}
\begin{threeparttable}
\begin{tabular}{cccccccc | c}
\hline
\multirow{2}{*}{Category} & \multirow{2}{*}{Dataset} & \multicolumn{6}{c}{Scanner}     & \multirow{2}{*}{Total} \\
&& m0 & m2 & m13 & m14 & m18 & m19 &                        \\ 
\hline
\multirow{4}{*}{ADNI-2}   & sMCI                     & 15 & 8  & 11  & 4   & 35  & 29  & 102                    \\
                          & pMCI                     & 9  & 14 & 6   & 7   & 22  & 9   & 67                     \\
                          & NC                       & 22 & 15 & 31  & 13  & 73  & 15  & 169                    \\
                          & AD                       & 22 & 19 & 22  & 12  & 43  & 10  & 128                    \\
\hline
\multicolumn{2}{c}{Total}                            & 68 & 56 & 70  & 36  & 173 & 63  & 466 \\
\hline
\end{tabular}
\end{threeparttable}
\end{table}

\subsection{The effect of the scanner on the predicted results}\label{s10}
\vspace{10pt}

\begin{table*}[ht!]
\centering
\caption{Results of MCI conversion prediction on the scanner-corrected ADNI-2 dataset.}
\label{tabs15}
\setlength{\tabcolsep}{3pt}
\begin{threeparttable}
\begin{tabular}{ccccccc}
\hline
Method                           & Student       & AT            & SP            & M3ID (Ours)   & M3ID+DML (Ours) & Teacher       \\
\hline
AUC without scanner   correction & 0.828 (0.006) & 0.846 (0.006) & 0.856 (0.007) & 0.866 (0.010) & 0.872 (0.003)   & 0.924 (0.004) \\
AUC with scanner   correction    & 0.822 (0.005) & 0.845 (0.011) & 0.850 (0.009) & 0.865 (0.006) & 0.866 (0.003)   & 0.924 (0.002) \\
\hline
\end{tabular}
\end{threeparttable}
\end{table*}

\begin{table*}[ht!]
\centering
\caption{Results of MCI conversion prediction on the scanner-corrected ADNI-1 dataset.}
\label{tabs16}
\setlength{\tabcolsep}{3pt}
\begin{threeparttable}
\begin{tabular}{ccccccc}
\hline
Method                         & Student         & AT              & SP            & M3ID (Ours)   & M3ID+DML (Ours) & Teacher       \\
\hline
AUC without scanner correction & 0.785   (0.005) & 0.789   (0.005) & 0.805 (0.006) & 0.812 (0.003) & 0.816 (0.009)   & 0.861 (0.008) \\
AUC with scanner correction    & 0.781   (0.004) & 0.788   (0.005) & 0.798 (0.007) & 0.804 (0.002) & 0.812 (0.012)   & 0.860 (0.009) \\
\hline
\end{tabular}
\end{threeparttable}
\end{table*}
\vspace{-10pt}

The MRI images of our studied ADNI subjects are collected using 20 different models of scanners (see Table S12). And the numbers of images scanned by each model of scanner differs dramatically. It is very difficult to investigate the effect of scanner on predicted results.

Following previous studies \citep{Dukart2011,Lin2018a} that perform age correction, we analogously perform scanner correction and investigate whether the prediction performance is affected after the correction. Specifically, we estimate the scanner-effect by fitting a linear regression model $\boldsymbol{y}_l=\boldsymbol{\omega}_l \boldsymbol{A} + \boldsymbol{b}_l$ at \textit{l}-th voxel: $\boldsymbol{y}_l \in \mathbb{R}^{1 \times N}$ is the vector of intensity values of N subjects at \textit{l}-th voxel; $\boldsymbol{\omega}_l \in \mathbb{R}^{1 \times C}$ is the learnable vector that weights scanners; $\boldsymbol{A} \in \mathbb{R}^{C \times N}$ is the matrix of the scanners of N subjects, and each subject is represented with a C-dimensional one-hot vector (C is the number of the models of scanners); and $\boldsymbol{b}_l \in \mathbb{R}^{1 \times N}$ is the vector of bias values of N subjects at \textit{l}-th voxel. For \textit{i}-th subject, the new intensity of \textit{l}-th voxel can be calculated as $y_l^{i'}=\boldsymbol{\omega}_l (\boldsymbol{\alpha}^C-\boldsymbol{\alpha}^i)+y_l^i$, where $y_l^i$ is original intensity, $\boldsymbol{\alpha}^i \in \mathbb{R}^{C \times 1}$ is the scanner vector of \textit{i}-th subject, and $\boldsymbol{\alpha}^C$ is a constant vector of $[1/C,1/C,…,1/C]^T \in \mathbb{R}^{C \times 1}$.

Considering that some models of scanners only have scanned limited numbers of images (see Table \ref{tabs12}), we choose 6 models of scanners that have top-6 numbers of images (see Table \ref{tabs13}-\ref{tabs14}). We trained the scanner correction model using the images of top-6 scanners in the training sets and applied it to the images of top-6 scanners in the test sets. We evaluated the trained deep learning models with the scanner-corrected images in the test sets. The deep learning models were the same as the models used for evaluations in Table \ref{tab2}-\ref{tab3}. The results are reported in Table \ref{tabs15}-\ref{tabs16}. The AUCs with scanner-correction are a bit lower than the AUCs without scanner-correction, but there are not statistically significant differences (t-test, $p > 0.05$). This suggests our predicted results are robust to effect of the scanner.

\subsection{Initially selected subjects regardless of MRI pre-processing}\label{s11}
We provide the numbers of subjects initially selected regardless of the MRI pre-processing in Table \ref{tabs17}. In summary, 4 images in ADNI-1 (1 sMCI, 2 pMCI, 1 NC) and 4 images in ADNI-2 (3 pMCI, 1 AD) did not pass the pre-processing. As for AIBL dataset, the number of initially selected subjects equals to the number of images passed pre-processing.

\begin{table}[t!]
\centering
\caption{Initially selected subjects regardless of MRI pre-processing.}
\label{tabs17}
\setlength{\tabcolsep}{3pt}
\begin{threeparttable}
\begin{tabular}{
p{30pt}<{\centering} p{35pt}<{\centering} p{35pt}<{\centering} p{50pt}<{\centering} p{60pt}<{\centering}
}
\hline
Dataset & Category & No. of subjects & Age Range & Females/Males \\
\hline
\multirow{4}{*}{ADNI-1} & sMCI & 101 & 56.1-87.9 & 34/67   \\
                        & pMCI & 166 & 55.2-88.3 & 66/100  \\
                        & NC   & 227 & 59.9-89.6 & 109/118 \\
                        & AD   & 179 & 55.1-90.9 & 87/92   \\
\hline
\multirow{4}{*}{ADNI-2} & sMCI & 115 & 55.0-88.6 & 48/67   \\
                        & pMCI & 79  & 56.5-85.9 & 36/43   \\
                        & NC   & 186 & 56.2-89.0 & 96/90   \\
                        & AD   & 141 & 55.6-90.3 & 56/85   \\
\hline
\end{tabular}
\end{threeparttable}
\end{table}

\subsection{Visualization of sMCI patch scores}\label{s12}

\begin{figure*}[b!]
\centerline{\includegraphics[width=1.5\columnwidth]{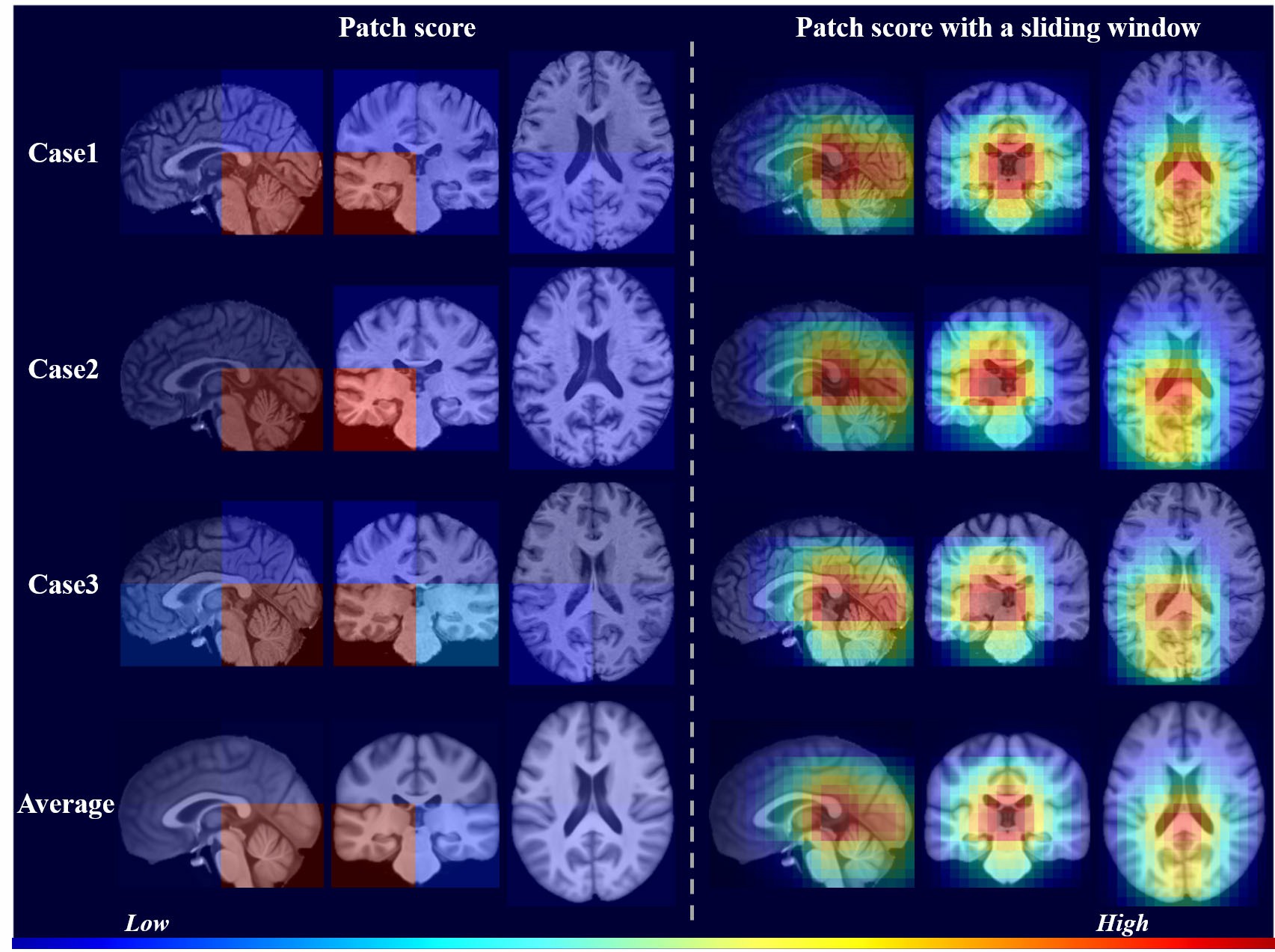}}
\setlength{\belowcaptionskip}{-10pt}
\caption{Visualization of patch scores of sMCI subjects. Left panel: 8 non-overlapping patches evenly divided from a whole MRI; right panel: patches extracted with a sliding window running across the whole MRI. All of the patches are of size 72$\times$88$\times$72. The average results are computed using all correctly predicted test cases.}
\label{figs6}
\end{figure*}

We presented the sMCI visualization with slices best showing the highlighted areas of sMCI in Fig. \ref{figs6}. By comparing Fig. \ref{fig6} with Fig. \ref{figs6}, we can see that the network focus on different areas for pMCI and sMCI. In sMCI brains, the highlighted areas include ventricle, corpus callosum, and posterior thalamus (Fig. \ref{figs6}). These areas are affected by AD and show differences between subjects with probable AD and those with memory impairment \citep{Jong2008,Tang2014,Elahi2015}. Fig. \ref{figs6} suggests that our methods guide the network to leverage the patch relationships and develop reasonable ability of localization.

\subsection{Visualizations of incorrectly predicted cases}\label{s13}
\begin{figure*}[ht!]
\centerline{\includegraphics[width=2.2\columnwidth]{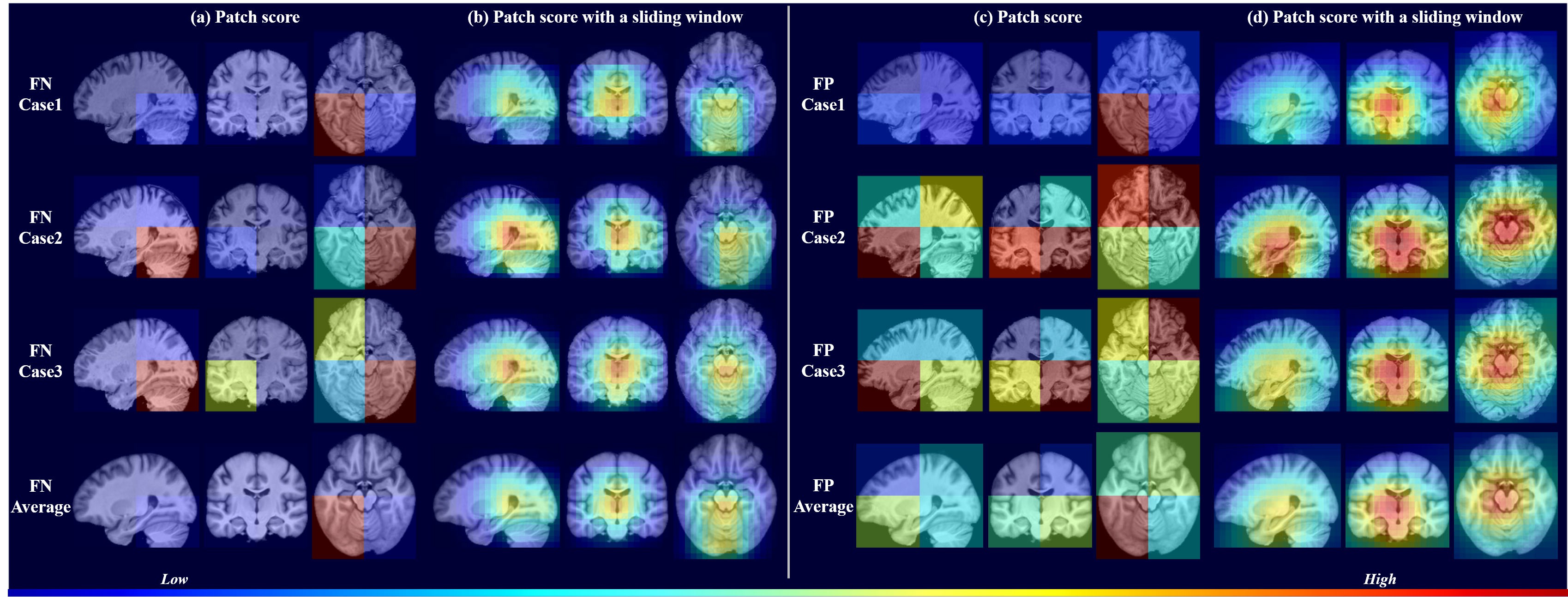}}
\setlength{\belowcaptionskip}{-10pt}
\caption{Visualization of patch scores of incorrectly predicted pMCI and sMCI subjects. FN refers to false negative subjects who have pMCI labels but are incorrectly predicted as sMCI. FP refers to false positive subjects who have sMCI labels but are incorrectly predicted as pMCI. Panel (a) and (c): 8 non-overlapping patches evenly divided from a whole MRI; panel (b) and (d): patches extracted with a sliding window running across the whole MRI. All of the patches are of size 72$\times$88$\times$72.}
\label{figs8}
\end{figure*}

We provide the visualizations of incorrectly predicted cases in Fig. \ref{figs8}. Compared with Fig. \ref{fig6}, the heatmaps of incorrectly predicted cases in Fig. \ref{figs8} are diffuse and less concentrated around the disease-related areas. In addition, the heatmaps of false negatives (FN) in Fig. \ref{figs8} are similar to the heatmaps of correctly predicted sMCI in Fig. \ref{fig6}. And the heatmaps of false positives (FP) in Fig. \ref{figs8} are similar to the heatmaps of correctly predicted pMCI in Fig. \ref{fig6}. This suggests that the network mistakes the cases by focusing on the areas related to the opposite class.



\end{document}